\documentclass[twocolumn]{aastex63}

\usepackage{textcomp}
\usepackage{bm,color}
\usepackage{verbatim}
\usepackage{amsmath}
\usepackage{ulem}
\usepackage{hyperref}  % <- new
\usepackage{booktabs}
\usepackage{bigstrut}
\usepackage{xspace}
\usepackage{multirow}
\usepackage{amsmath}
\usepackage{graphicx}
\usepackage{threeparttable}

\def \BE{\begin{equation}}
\def \EE{\end{equation}}	
\def \BC{\begin{center}}
\def \EC{\end{center}}
\def \BEA{\begin{eqnarray}}
\def \EEA{\end{eqnarray}}

\def \SIGMA8{\sigma_{8}}

\def\lcdm{\ensuremath{\Lambda {\rm CDM}}}

\begin{document}

%\maketitle
\title[Deep learning: Galaxy cluster mass estimates]{Mass Estimation of Galaxy Clusters with Deep Learning I: Sunyaev-Zel'dovich Effect }

\shortauthors{N.~Gupta, et al.}
\author[0000-0001-7652-9451]{N.~Gupta} \thanks{nikhel.gupta@unimelb.edu.au} \affiliation{School of Physics, University of Melbourne, Parkville, VIC 3010, Australia} 
\author[0000-0003-2226-9169]{C.~L.~Reichardt} \affiliation{School of Physics, University of Melbourne, Parkville, VIC 3010, Australia}

\begin{abstract}
We present a new application of deep learning to infer the masses of galaxy clusters directly from images of the microwave sky. 
Effectively, this is a novel approach to determining the scaling relation between a cluster's Sunyaev-Zel'dovich (SZ) effect signal and mass. 
The deep learning algorithm used is mResUNet, which is a modified feed-forward deep learning algorithm that broadly combines residual learning, convolution layers with different dilation rates, image regression activation and a U-Net framework. 
We train and test the deep learning model using simulated images of the microwave sky that include signals from the cosmic microwave background (CMB), dusty and radio galaxies, instrumental noise as well as the cluster's own SZ signal. 
The simulated cluster sample covers the mass range 1$\times 10^{14}~\rm M_{\odot}$ $<M_{200\rm c}<$ 8$\times 10^{14}~\rm M_{\odot}$ at $z=0.7$. 
The trained model estimates the cluster masses with a 1\,$\sigma$ uncertainty $\Delta M/M \leq 0.2$, consistent with the input scatter on the SZ signal of 20\%. 
We verify that the model works for realistic SZ profiles even when trained on azimuthally symmetric SZ profiles by using the Magneticum hydrodynamical simulations. 
\end{abstract}

\keywords{cosmic background radiation - large-scale structure of universe - galaxies: clusters: general}

\section{Introduction} 
\label{sec:intro}
Galaxy clusters reside in the most massive gravitationally bound halos in the cosmic web of large scale structure (LSS) and can be observed across the electromagnetic spectrum. 
In recent years, the Sunyaev-Zel'dovich (SZ) effect \citep[][]{sunyaev70, sunyaev72}, the inverse-Compton scattering of the cosmic microwave background (CMB) photons by the energetic electrons in the intracluster medium, has emerged as a powerful tool to detect galaxy clusters in the millimetre wavelength sky. 
Since  \citet{staniszewski09} presented the first SZ-discovered clusters, the South Pole Telescope \citep[SPT;][]{carlstrom11}, the Atacama Cosmology Telescope \citep[ACT;][]{fowler07} and the {\it Planck} satellite \citep{planck06} have released catalogs of hundreds to thousands of newly discovered clusters \citep[e.g.][]{planck16-24,hilton18, huang19, bleem19}. 
These cluster samples are significant because the abundance of galaxy clusters is one of the most promising avenues to constrain different cosmological models \citep[e.g.][]{mantz08, vikhlinin09, hasselfield13, planck16-24, dehaan16, bocquet19}.

With ongoing \citep[e.g. SPT-3G, AdvancedACT][]{benson14, henderson16} and upcoming \citep[e.g. Simons Observatory, CMB-S4][]{simons19, cmbs4-19} CMB surveys, we expect to detect $>$10$^4$ galaxy clusters. 
These cluster samples could have a ground-breaking impact on our understanding of the expansion history and structure growth in the universe, but only if we can improve the calibration of cluster masses \citep[see, e.g.][]{bocquet15, Planck15}.

Observationally, several techniques have been used to measure the masses of galaxy clusters, such as optical weak lensing \citep[e.g.][]{johnston07, gruen14, hoekstra15, stern18, mclintock19}, CMB lensing \citep[e.g.][]{baxter15, madhavacheril15, planck16-24, raghunathan19}, and dynamical mass measurements  \citep[e.g.][]{biviano13,sifon16,capasso19}. 
These techniques are typically used to calibrate the scaling relationship between mass and an easily-measurable observable such as the richness or SZ signal \citep[e.g.][]{sifon13, mantz16,stern18}. 
The latter is particularly interesting as numerical simulations have shown that the integrated SZ signal is tightly correlated with the mass of clusters \citep[e.g.][]{lebrun17, gupta17b}.  

In recent years, deep learning has emerged as a powerful technique in computer vision.
In this work, we demonstrate the first use of a deep learning network to estimate the mass of galaxy clusters from a millimeter wavelength image of the cluster.  
We employ a modified version of a feed-forward deep learning algorithm, mResUNet that combines residual learning \citep{he15d} and U-Net framework \citep{ronneberger15d}.
We train the deep learning algorithm with a set of simulations that  include the cluster's SZ signal added to Gaussian random realizations of the CMB, astrophysical foregrounds, and instrumental noise. 
We use the trained mResUNet model to infer the mass from a test data set, which is not used in the training process.
We also test the accuracy of the trained model using hydrodynamical simulations of galaxy clusters, which again are not used in the training process.

The paper is structured as follows. In Section~\ref{sec:methods}, we describe the deep learning reconstruction model and the microwave sky simulation data. 
In Section~\ref{sec:optimisation}, we describe the optimization process and the relevant hyper-parameters of the deep learning model.
In Section~\ref{sec:results}, we present mass estimations using the images from test data sets as well as the images from the external hydrodynamical simulations of SZ clusters.
Finally, in Section~\ref{sec:conclusions}, we summarize our findings and discuss future prospects.

Throughout this paper, $M_{\rm 200c}$ is defined as the mass of the cluster within the region where the average mass density is 200 times the critical density of universe. 
The central mass and the $1\,\sigma$ uncertainty is calculated as median and half of the difference between the $16^{\rm th}$ and $84^{\rm th}$ percentile mass, respectively.

\section{Methods}
\label{sec:methods}
In this section, we first describe the deep learning algorithm, and then present the microwave sky simulations that are used to train and test the deep learning model. 

\begin{figure*}
%\centering
\includegraphics[trim={2.5cm 3.5cm 2.cm 6.cm},clip,width=18cm, scale=0.5]{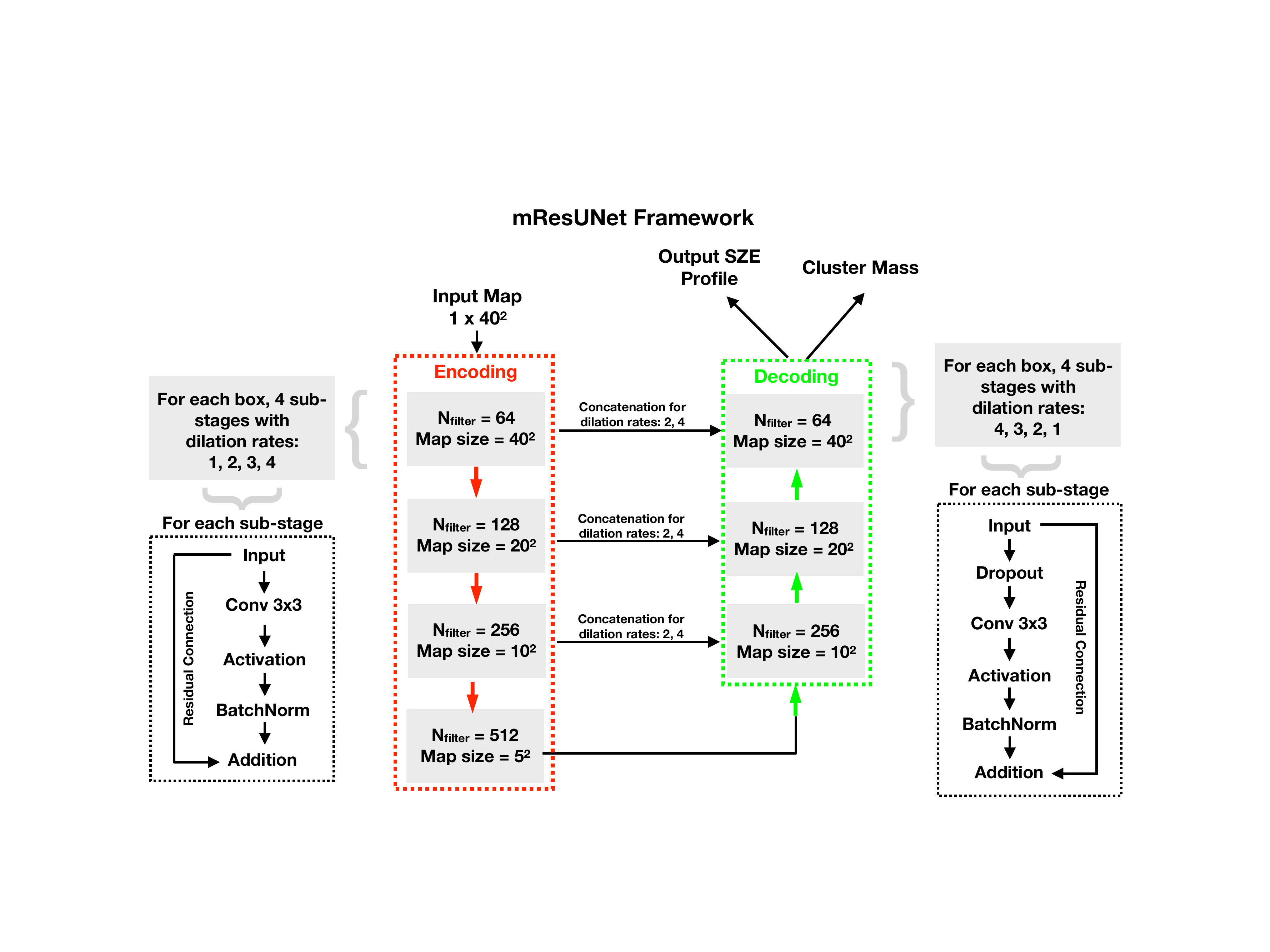}
\caption{The mResUNet framework with decoding (red dashed box) and encoding phases (green dashed box). 
Each gray coloured box in these phases represents a convolution block.
We change the number of filters and the map size by down sampling (red arrows) and up sampling (green arrows) the feature maps in the encoding and the decoding phases, respectively.
The convolution block has four sub-stages where convolution operations are applied with different dilation rates of N = 1, 2, 3 and 4.
All sub-stages have convolution, activation and batch normalization layers, and residual connections are applied between the input and output feature maps.
The sub-stages of convolution blocks in decoding phase have an extra dropout layer to prevent model over-fitting.
Skip connections are used to concatenate feature maps from the encoding convolution blocks to corresponding blocks in decoding phase that helps in retrieving the lost spatial information due to down sampling (see Section~\ref{sec:deep_learning}).} 
\label{FIG:ResUNet}
\end{figure*}
\subsection{Deep Learning Model}
\label{sec:deep_learning}
In recent years, deep learning algorithms have been extensively used in range of astrophysical and cosmological problems \citep[e.g.][]{george18d, mathuriya18d, allen19d, bottrell19d, alexander19d, fluri19d}. 
Recent studies have applied deep learning \citep[][]{ntampaka19a, ho19} and machine learning \citep[e.g.][]{ntampaka15,armitage19,green19} algorithms to estimate galaxy cluster masses using mock X-ray and velocity dispersion  observations. These studies found that these techniques produce more accurate X-ray and dynamical mass estimates than conventional methods.

In this work, we apply the mResUNet algorithm to extract the SZ profiles and the cluster masses from the simulated microwave sky maps. 
ResUNet is a feed-forward deep learning algorithm that was first introduced for segmentation of medical images \citep{kayalibay17} and to extract roads from maps \citep{zhang18d}, and later applied to a number of problems. 
The original algorithm was modified by \citet{caldeira19d} to do image to image regression, i.e. get an output image that is a continous function of the input image. 
We implement further modifications to the network to extract small and large scale features in the map.
This modified ResUNet, or mResUNet, algorithm is well suited to astrophysical problems, such as the current use case of estimating the SZ signal from an image of the sky. 

The mResUNet is a convolutional neural network and its basic building block is a convolution layer which performs discrete convolutions \citep[see][for a recent review]{gu15d}. 
The aim of the convolution layer is to learn features of an input map.  
Convolutional neural networks assume that nearby pixels are more strongly correlated than the distant ones.
The features of nearby pixels are extracted using filters that are applied to a set of neighbouring pixels.
This set of neighbouring pixels is also called the receptive field. 
The filter applied to a set of pixels is typically a $k \times k$ array with $k=1,3,5, ...$, and the size of the filter ($k \times k$) is denoted as the kernel size.
A filter with a given kernel-size is moved across the image from top left to bottom right and at each point in the image a convolution operation is performed to generate an output.
Several such filters are used in a convolution layer to extract information about different aspects of the input image.
For instance, one filter can be associated to the central region of the galaxy cluster and rest of the filters could extract information from the other parts of cluster. 
The filters can extract information across different length scales by using different dilation rates instead of increasing the kernel size. 
A dilation rate of N stretches the receptive field by $k + (k-1)(N-1)$, thus doubling the dilation rate will increase the receptive field to $5 \times 5$ for $k$=3.
These dilated convolutions systematically aggregate multi-scale contextual information without losing resolution \citep{yu15d}. 
 
The total receptive field increases for each pixel of the input image as we stack several convolution layers in the network.
An activation function is applied after each convolution layer, which is desirable to detect non-linear features and results into a highly non-linear reconstruction of input image \citep[see][for a recent review]{nwankpa18d}. 
Each convolution layer produces a feature map for a given input image.
The feature map ($f_l$) for a convolution layer ($l$) is obtained by convolving the input from a previous layer ($x_{l-1}$) with a learned kernel, such that, the feature value at location ($i,j$) is written as
\BE
\centering
\label{EQ:feature}
f_l^{i,j} = w_l^T ~ x_{l-1}^{i,j} ~ + ~ b_l,
\EE
where $w_l$ is the weight vector and $b_l$ is the bias term. 
The weights are optimized using gradient descent \citep[e.g.][]{ruder16d} that involves back-propagation from the final output, back to each layer in reverse order to update the weights.

The mResUNet architecture used in this work has following main components.
\begin{enumerate}
\item 
We base our architecture on the encoder-decoder paradigm. 
This consists of a contracting path (encoder) to capture features, a symmetric expanding path (decoder) that enables precise localization and a bridge between these two. 
Figure~\ref{FIG:ResUNet} shows the full UNet framework, where the red and the green dashed lines point to encoding and decoding frameworks, respectively.
\item 
Each grey coloured box corresponds to a convolution block.
We increase the filter size from 64 to 512 and use strides \citep[e.g.][]{dumoulin16d} to reduce the size of feature map by half whenever filter size is doubled (red arrows) during the encoding phase of the network. 
This process is known as down sampling by striding.
For the decoding phase, we increase the size of feature map by up sampling (green arrows). 
Each convolution block has 4 sub-stages where convolution operations are applied with different dilation rates of N = 1, 2, 3 and 4, while keeping the stride length to unity, whenever dilation rate is not 1.
This improves the performance by identifying correlations between different locations in the image \citep[e.g.][]{yu15d,chen16d,chen17d}.
\item 
The feature maps from two sub-stages (dilation rates N=2, 4) of first three encoding convolution blocks are cross concatenated with the corresponding maps from decoding blocks using skip connections. 
These connections are useful to retrieve the spatial information lost due to striding operations \citep[e.g.][]{drozdzal16d}.
\item
Each sub-stage of encoding and decoding convolution blocks has fixed number of layers. 
Among these the convolution, the activation and the batch normalization layers are present in all sub-stages.
The batch normalization layer which is helpful in improving the speed, stability and performance of the network \citep{Ioffe15d}.
The input to these layers is always added to its output, as shown by the connection between input and addition layers. 
Such connections are called residual connections \citep{he15d} and they are known to improve the performance of the network \citep[e.g.][]{zhang18d, caldeira19d}.
\item 
A large feed-forward neural network when trained on a small set of data, typically performs poorly on the test data due to over-fitting. 
This problem can be reduced by randomly omitting some of the features during the training phase by adding dropout layers to the network \citep{hinton12d}.
We add dropout layers to the decoding phase of the network.
\end{enumerate}

\begin{figure*}
\centering
\includegraphics[trim={0 8.5cm 0 5.5cm},clip,width=18cm, scale=0.5]{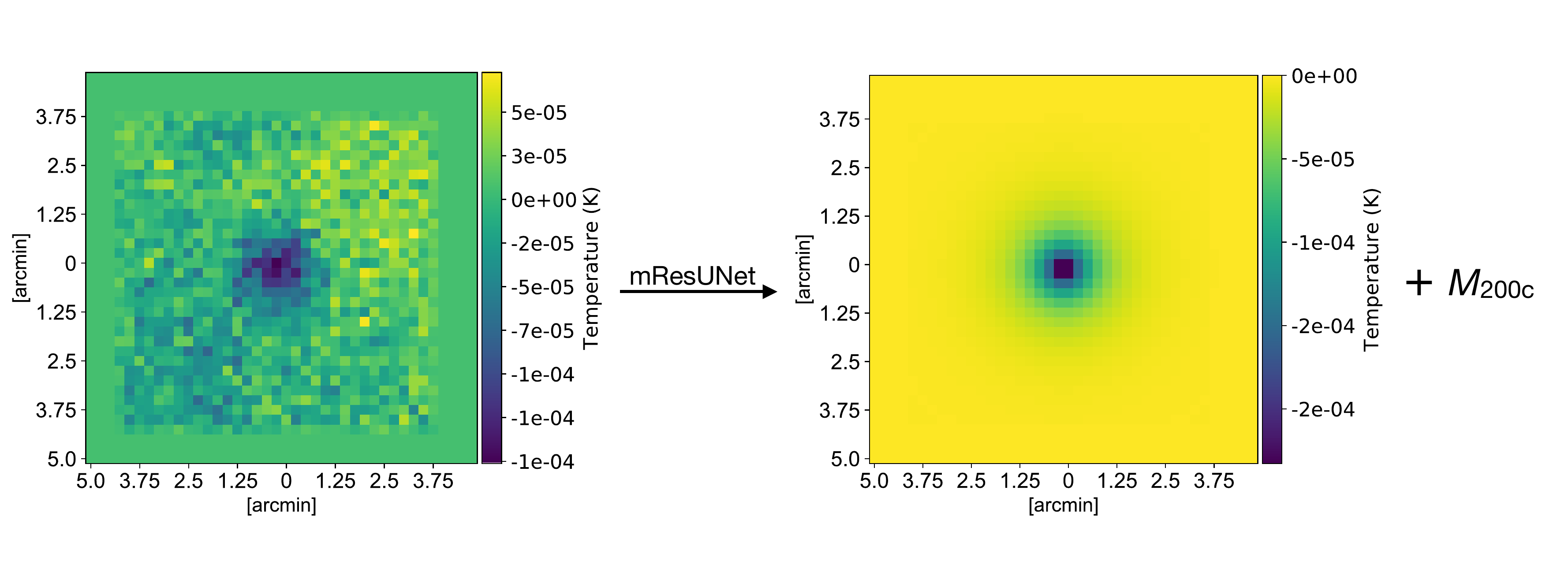}
\caption{The work flow from simulations to mass estimations: Left panel shows an example of the microwave sky CMB map with the SZ imprint of a cluster with $M_{\rm 200c} = 5\times 10^{14}~\rm M_{\odot}$ at $z=0.7$. This map includes $5~\rm \mu K$-arcmin white noise, foreground power estimates from \citet{george15} and is smoothed by a $1^{\prime}$ beam. Several such maps for different cluster masses are used for training and validation of the neural network. Right panel shows SZ profile computed using best fit GNFW profile and mass-observable scaling relation in \citet{arnaud10}. In addition to microwave sky maps, the training set includes the true SZ profiles and the true mass of clusters as labels to train the model. A different set of simulations are created for testing the model and the trained model is then used to predict the SZ profiles and the mass of clusters directly from the CMB maps of testing set.} 
\label{FIG:simulation}
\end{figure*}
\subsection{Microwave Sky Simulations}
\label{sec:SZE}
In this section, we describe the microwave sky simulations of SZ clusters. 
We create 19 distinct set of simulations for galaxy clusters with $M_{200\rm c} =$ (0.5, 0.75, 1, 1.5, 2, 2.5, 3, 3.5, 4, 4.5, 5, 5.5, 6, 6.5, 7, 7.5, 8, 9, 10)$\times 10^{14}~\rm M_{\odot}$ at $z=0.7$.
For each mass, we create 800 simulated $10^{\prime} \times 10^{\prime}$ sky images, centered on the cluster with a pixel resolution of $0.25^{\prime}$. 
While upcoming CMB surveys (see Section~\ref{sec:intro}) will observe the microwave sky at multiple frequencies, we make the simplifying assumption in this work to focus on single-frequency maps at 150\,GHz. 
The sky images include realisations of the CMB, white noise, SZ effect, cosmic infrared background (CIB) and radio galaxies.
The CMB power spectrum is taken to be the lensed CMB power spectrum calculated by \texttt{CAMB}\footnote{\url{https://camb.info/}} \citep{lewis00} for the best-fit {\it Planck} \lcdm{} parameters  \citep{planck18-1}. 
The foreground terms, the thermal and kinematic SZ effect from unrelated halos, cosmic infrared background (CIB) and radio galaxies, are taken from \citet{george15}.  
We assume the instrumental noise is white with a level of 5 $\rm \mu K$-arcmin, similar to what was achieved by the SPTpol survey \citep{henning18}.
Note that these simulations neglect non-Gaussianity in the astrophysical foregrounds, as well as gravitational lensing of the CMB by large-scale structure besides the cluster itself. 
Future work should assess the impact of these sources of non-Gaussianity on the deep learning estimator.

We assume the cluster's own SZ signal follows the Generalized Navarro-Frenk-White \citep[GNFW;][]{nagai07} pressure profile, with parameters as a function of mass and redshift taken from the best-fit values in \citet{arnaud10}. 
In addition unless noted, we add a 20\% log-normal scatter on the modelled amplitude of the SZ signal.
This is slightly larger than the amount of scatter ($\sigma_{{\rm ln} Y} \sim 0.16$) found in the calibration of scaling relations using a light cone from large hydrodynamical simulations \citep[e.g.][]{gupta17b}, and thus conservative.

We convolve these maps with $1^{\prime}$ Gaussian beam which is consistent with ground based SPT and ACT experiments at 150~GHz, and apply apodization. 
One of these cluster cutouts is shown in Figure~\ref{FIG:simulation}  for $M_{200 \rm c} = 5\times 10^{14}~\rm M_{\odot}$ and a random CMB realisation.
In addition to these microwave sky SZ cluster maps, we save the corresponding SZ profiles and the mass of clusters that are used as labels in the training process.
In order to recover masses from a framework designed to recover images, we set the central pixel value of the `mass map' to be proportional to the cluster mass. 
We then extract this central pixel value when reporting the recovered mass constraints.

\subsection{Uncertainties in SZ-Mass Scaling Relation}
\label{sec:systematics}
The deep learning model in this work is trained on a specific SZ-mass scaling relation, here chosen to be the Arnaud model. 
Of course, we have imperfect knowledge of the relationship between a typical cluster's SZ flux and mass. 
Recent measurements of the SZ-mass scaling relation are uncertain at the O(20\%) level \citep{dietrich19, bocquet19}. 
This uncertainty is a fundamental limit to how well methods like this one that estimate cluster masses from the SZ signal can perform.
However, this uncertainty can be reduced by calibrating the relationship on samples of clusters using weak gravitational lensing  \citep[e.g.][]{corless09,becker11}. 
Several programs employing gravitational lensing are currently underway \citep[e.g. Dark Energy Survey, Hyper Suprime-Cam Survey][]{mclintock19, murata19} or expected to start observing in a near future \citep[e.g. LSST, {\it Euclid}][]{lsst09,laurejis11}, will lead to much tighter constraints on the the SZ-mass scaling relation.
In this paper, we test the deep learning model on the simulated sky maps with SZ profiles taken from the Arnaud scaling relation and from the hydrodynamical simulations with a different intrinsic SZ-mass scaling relation.

\section{Training and Optimisation}
\label{sec:optimisation}
The mResUNet model described in Section~\ref{sec:deep_learning} and Figure~\ref{FIG:ResUNet} takes images as input and outputs same sized images after passing through several convolutional blocks.
This process is repeated for a number of epochs, where one epoch is when entire training data are passed through the neural network once.
The data are divided into three parts: training, validation and test sets.

The training dataset includes images of the microwave sky simulations of SZ clusters, the corresponding true SZ profiles and the true mass of clusters. 
As described in Section~\ref{sec:SZE}, both CMB maps and SZ profiles have a characteristic 20\% log-normal SZ-mass scatter and all CMB maps have Gaussian random realizations of CMB.
To make these simulations more realistic, we add foregrounds, 5 $\rm \mu K$-arcmin white noise and $1^{\prime}$ beam smoothing to these maps.
We normalize all maps, so that, the minimum and maximum pixel value is between -1 and 1, respectively, to improve the performance of network.
This is done by dividing the image pixels by a constant factor across all cluster masses.
Our training data has 400 maps for each cluster and corresponding labels (true SZ profiles and true mass of clusters).
For training, we only take cluster simulations with $M_{200\rm c} =$ (1, 2, 3, 4, 5, 6, 7, 8)$\times 10^{14}~\rm M_{\odot}$ and leave others for testing the model. 

The validation set has same properties as the training set and is also used in the training phase to validate the model after each epoch.
This is helpful as a non-linear model is more likely to get high accuracy and over-fit when trained with training data only. 
Such a model gives poor performance with the test data.
The validation of the model after every epoch ensures regular checks on model over-fitting and is useful to tune the model weights.
We use 200 maps for each cluster mass and corresponding labels as our validation data. 

The test datasets are never used in the training phase and are kept separately to analyse the trained model.
We keep 200 CMB temperature maps and corresponding labels for testing.
In addition to the cluster $M_{200\rm c}$ used in training, we test our model for cluster masses that were not the part of training or validation process ,that is, clusters with $M_{200\rm c} =$ (0.5, 0.75, 1.5, 2.5, 3.5, 4.5, 5.5, 6.5, 7.5, 9, 10)$\times 10^{14}~\rm M_{\odot}$.

The maps from the training set are passed through the neural networks with a batch size of 4 and a training loss is computed as mean-squared-error (MSE) between the predicted and the true labels after each batch.
Batch after batch, the weights of the network are updated using the gradient descent and the back-propagation (see Section~\ref{sec:deep_learning}).
In this work, we use Adam optimizer \citep[an algorithm for first-order gradient-based optimization, see][]{kingma14d} with an initial learning rate of 0.001. 
After each epoch, the validation loss (or validation MSE) is calculated and we change the learning rate by implementing callbacks during the training, such that, the learning rate is reduced to half if the validation loss does not improve for five consecutive epochs.
In addition, to avoid over-fitting, we set a dropout rate of 0.3 in the encoding phase of the network.
We consider the network to be trained and stop the training process, if the validation loss does not improve for fifteen epochs.

Every convolution block in encoding, bridging and decoding phase has a convolution layer, an activation layer and a batch normalization layer.
The kernel-size of each convolution layer is set to $3\times 3$ and we change stride length from 1 to 2, whenever filter size is doubled.
All activation layers in the network have Scale Exponential Linear Unit \citep[SELU][]{klambauer17d} activation functions which induce sellf-normalizing properties, such that, activations close to zero mean and unit variance converge towards zero mean and unit variance, when propagated through many network layers, even under the presence of noise and perturbations.
Only for the final layer, linear (or identity) activation function is used to get same sized output images as inputs.
The network has approximately 16 million parameters and is trained on a single GPU using Keras with a TensorFlow backend.
\begin{figure*}
\centering
\includegraphics[width=18cm, scale=0.5]{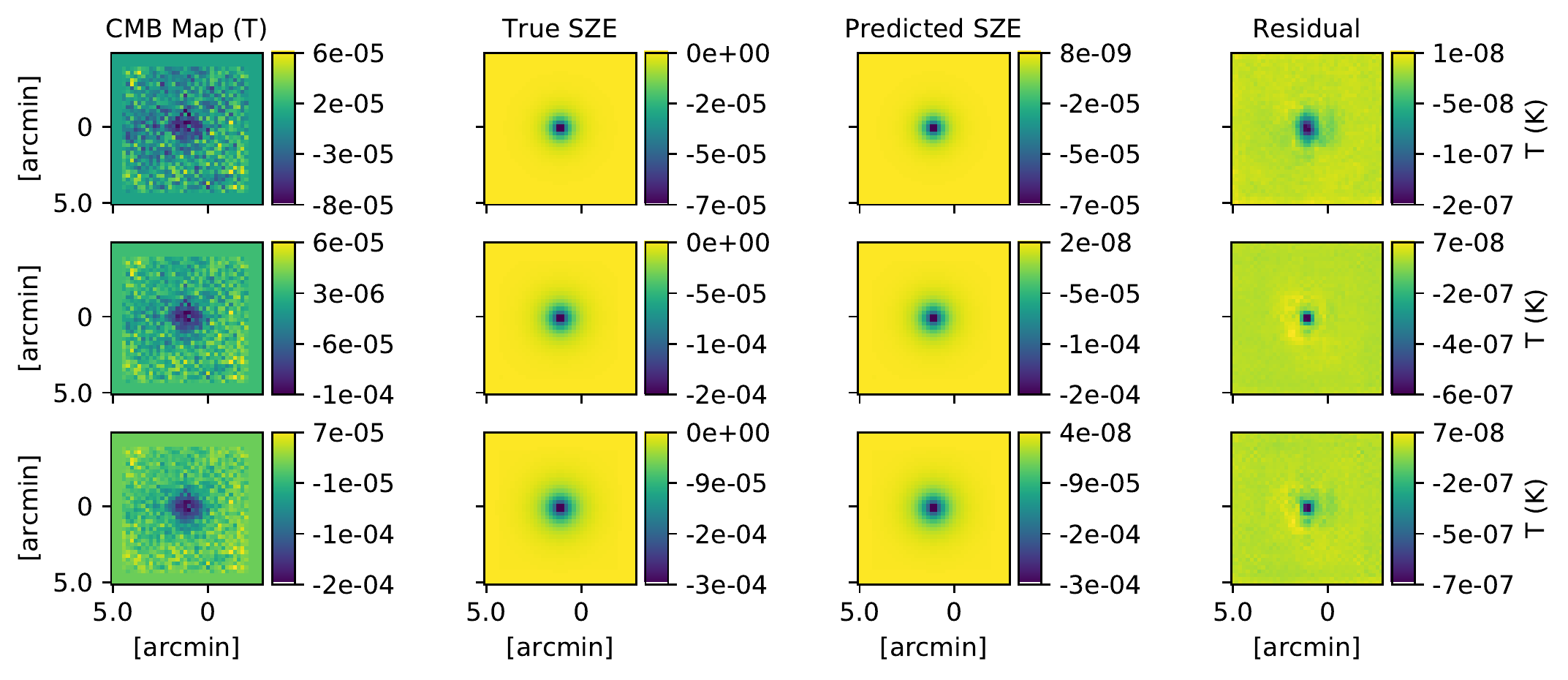}
\caption{SZ profile predictions: Examples of CMB temperature maps (column 1), true and predicted mean SZ profiles (columns 2 and 3, respectively) and residual between true and predicted mean SZ profiles (column 4). From top to bottom, these maps indicate different clusters with $M_{200\rm c} =$ (2, 4, 6)$\times 10^{14}~\rm M_{\odot}$. The difference between the true and predicted profiles  is small, such that, the residuals are at-least two order of magnitude smaller than the true SZ signal. This demonstrates high accuracy in the image-to image reconstruction ability of our trained model.} 
\label{FIG:SZEprofile}
\end{figure*}
\begin{figure}
\includegraphics[width=8.4cm, scale=0.4]{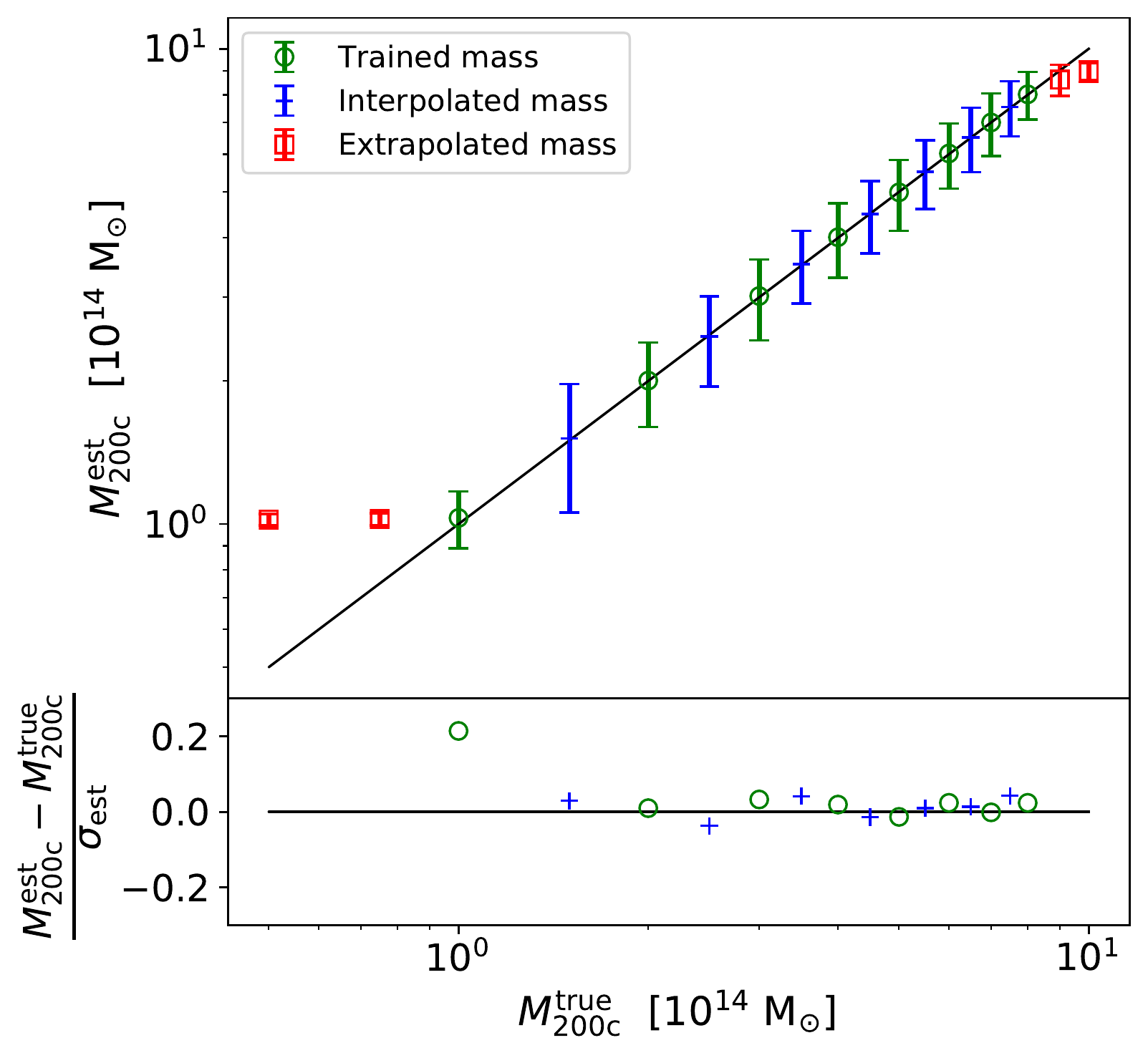}
\caption{The trained model returns unbiased mass estimates for masses within the training range. 
For lower (higher) masses, the estimated mass plateaus at the lowest (highest) mass in the training set.  
The top panel plots the estimated versus true mass of clusters using a test data set of 200 CMB temperature maps per cluster mass.
The points in red show the results for clusters with masses outside the trained range (extrapolation). 
The green points show the results for clusters with masses equal to one of the training sets. 
The blue points show the results for clusters with masses between the trained masses (interpolation). 
The  bottom panel shows the significance of the difference between the estimated and true masses for each set. 
The bias increases for masses at the edge of the trained range, but is always much less than $1\,\sigma$. 
}
\label{FIG:mass}
\end{figure}
\section{Results}
\label{sec:results}

We now look at the performance of the trained deep learning model on the test data. 
We test the performance of the trained model in three regimes: (i) cluster masses within the trained mass range (i.e. interpolation); (ii) cluster masses outside the trained mass range (i.e. extrapolation); and (iii) more realistic SZ clusters drawn from a large hydrodynamical simulation, the  {\it Magneticum Pathfinder Simulation}\footnote{http://www.magneticum.org/} (MPS). 
We find the model performs well in the first and third cases, but fails in the extrapolation case. 

\subsection{Predictions with Trained Cluster Mass}
\label{sec:predictions1}
We use the test data having 200 CMB maps for each of the clusters with $M_{200\rm c} =$ (1, 2, 3, 4, 5, 6, 7, 8)$\times 10^{14}~\rm M_{\odot}$.
This testing mass is same as that used in the training of our mResUNet model.
These test maps are not used in training and validation phases and are distinct due to the Gaussian random realizations of the CMB and foregrounds as well as the 20\% log-normal scatter in the estimation of the SZ signal.
The trained model predicts SZ profiles as well as the mass of clusters from the CMB maps.
The first column in Figure~\ref{FIG:SZEprofile} shows examples of the input CMB temperature maps for clusters with $M_{200\rm c} =$ (2, 4, 6)$\times 10^{14}~\rm M_{\odot}$ from top to bottom.
The second and the third columns show true and predicted mean SZ profiles, respectively, for 200 test maps.
The last column shows residual signals, that is, the difference between the true and the predicted mean SZ profiles.
This demonstrates that the deep learning model reconstructs SZ profiles with a high accuracy, such that, the residual signal is atleast two-orders of magnitude smaller than the true SZ signal.

We simultaneously estimate the mass of galaxy clusters using the trained model. 
As described in Section~\ref{sec:SZE}, this is done by multiplying the central pixel of the predicted normalized NFW profiles by the mean mass of the training sample.
The top panel in Figure~\ref{FIG:mass} shows the estimated mass of clusters as a function of their true mass (green data points).
This demonstrates that our trained mResUNet model can estimate cluster masses with high accuracy.
For instance, we find $M_{200 \rm c}^{\rm est}=(1.99\pm 0.40) \times 10^{14}~\rm M_{\odot}$ for a cluster with $M_{200 \rm c}^{\rm true}=2 \times 10^{14}~\rm M_{\odot}$ and $\Delta M/M \leq 0.2$ for all cluster masses.
The bottom panel shows the ratio of the difference between estimated and the true mass of clusters to the estimated uncertainty.
This indicates that the mass estimations with our trained neural network model are consistent with the input mass at $1\,\sigma$ level.

\subsection{Predictions with Interpolated and Extrapolated Cluster Mass}
\label{sec:predictions2}
In this section, we present the mass estimations using the test maps for clusters with untrained masses.
We divide these samples into two types, that is, interpolated and extrapolated cluster masses.
The first type of  clusters lie with in the mass range of trained cluster sample with $M_{200\rm c} =$ (1.5, 2.5, 3.5, 4.5, 5.5, 6.5, 7.5)$\times 10^{14}~\rm M_{\odot}$ and the second type of clusters are out of the training mass range with $M_{200\rm c} =$ (0.5, 0.75, 9, 10)$\times 10^{14}~\rm M_{\odot}$.
As before, white noise and 20\% log-normal scatter is added to the SZ signal, and these maps are smoothed by a $1^{\prime}$ beam as well.

The top panel in Figure~\ref{FIG:mass} shows the estimated and the true mass for interpolated (blue) and extrapolated (red) test data sets. 
The bottom panel shows the ratio of the difference between estimated and true mass of clusters to the estimated uncertainty.
The $1\,\sigma$ error in the mass estimation for interpolated clusters is consistent with the true input mass.
The uncertainties are similar to those from trained sample (Section~\ref{sec:predictions1}), for instance, the $M_{200 \rm c}^{\rm est}=(3.52 \pm 0.61) \times 10^{14}~\rm M_{\odot}$ for a cluster with $M_{200 \rm c}^{\rm true}=3.5 \times 10^{14}~\rm M_{\odot}$.
The $\Delta M/M \leq 0.21$ for all cluster masses, except for the cluster with $M_{200 \rm c}^{\rm true}=1.5 \times 10^{14}~\rm M_{\odot}$ where $\Delta M/M = 0.3$.
This shows that our trained neural network can be used to make accurate mass estimations for all clusters inside the mass range of our training sample.
As expected, for extrapolated clusters, the neural network does not estimate correct masses. 
One exception is the cluster with $M_{200\rm c} =$ 9$\times 10^{14}~\rm M_{\odot}$ for which the extrapolation out of trained mass range gives consistent predictions. We consider this a random occurrence given the image to image regression framework of our model.
This indicates that the training sample needs to be expanded to accurately estimate the mass of clusters that are outside the range of our training sample.

\begin{figure}
\centering
\includegraphics[width=8.2cm, scale=0.5]{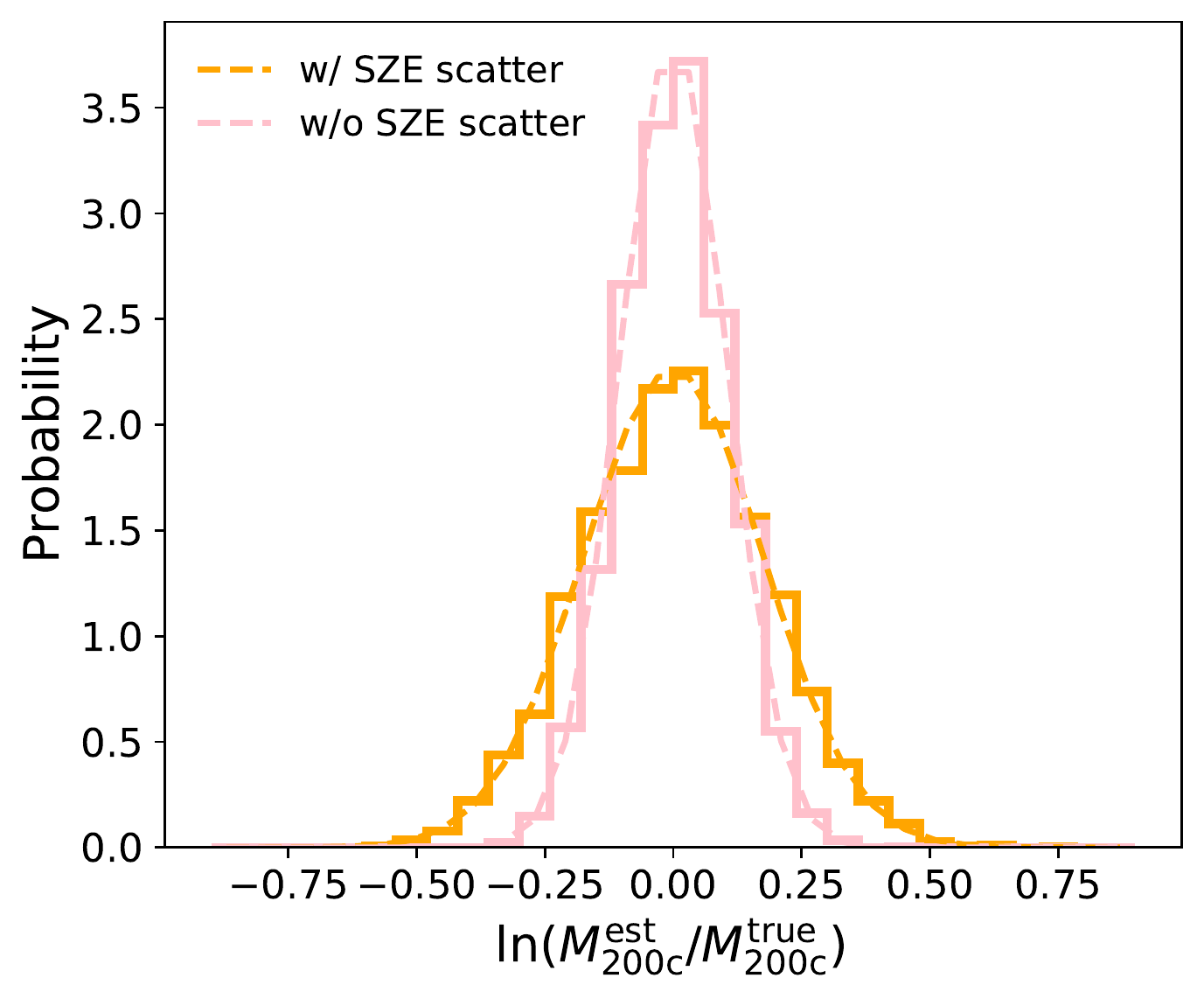}
\caption{
The scatter in the estimated mass is dominated by the input scatter in the SZ-mass relationship. 
This plot shows the difference in the log-normal masses,  $\ln(M^{\rm est}_{\rm 200c}/M^{\rm true}_{\rm 200c})$ for a set of 1000 clusters with masses drawn uniformly from the range 2$\times 10^{14}~\rm M_{\odot}$ $<M_{200\rm c}<$ 7$\times 10^{14}~\rm M_{\odot}$. 
The orange line shows the results when the test set includes a 0.2 log-normal scatter on the SZ signal, while the pink line shows the results with no scatter. 
The best-fit Gaussian (dashed lines) width in the two cases is $0.180\pm 0.013$ and $0.100\pm 0.012$ respectively. 
This shows that the dominant uncertainty in the model's mass estimate is due to the input SZ scatter in the simulations. 
} 
\label{FIG:mass_scatter}
\end{figure}
\subsection{Sources of uncertainty in the mass estimate}
\label{sec:scatter}

In evaluating the deep learning method's performance, an interesting question is what portion of the final mass uncertainty is due to the intrinsic scatter in the SZ signal between two clusters of the same mass as opposed to uncertainty in the measurement. 
We do this by creating two sets of 1000 test maps including the cluster SZ signal along with CMB, instrumental noise and foregrounds. 
The cluster masses are distributed across the training range 2$\times 10^{14}~\rm M_{\odot}<M_{200\rm c}< 7\times 10^{14}~\rm M_{\odot}$. 
In the first set, the cluster SZ signal is added with a 20\% log-normal scatter, while the second set has zero scatter. 
The training of mResUNet network is the same in both cases as detailed in Section~\ref{sec:deep_learning}.

Figure~\ref{FIG:mass_scatter} shows normalized histogram of the natural log of the ratios of estimated and true cluster masses, in orange for the simulations with 20\% scatter, and pink for the simulations with no scatter. 
We fit a Gaussian to each histogram to calculate the log-normal scatter, while using bootstrapping to estimate the error.
The observed log-normal scatter in the recovered mass is $0.180\pm 0.013$ for simulations with 20\% intrinsic SZ scatter, and $0.100\pm0.012$ for the no-scatter simulations. 
The apparent small reduction in scatter in the first case is consistent with a statistical fluctuation at $1.5\,\sigma$. 
These results clearly demonstrate that the deep learning method to estimate cluster masses from the SZ signal has reached the theoretical lower limit set by the intrinsic SZ scatter.

A secondary implication of this result is that although upcoming CMB surveys with multiple observing frequencies and lower noise levels will yield higher fidelity measurements of the cluster SZ signal, this improvement may not translate to better mass estimates. 
Nevertheless, we plan to consider the impact of multiple frequency maps on the deep learning analysis in future work. 

\begin{figure}
\centering
\includegraphics[width=8cm, scale=0.5]{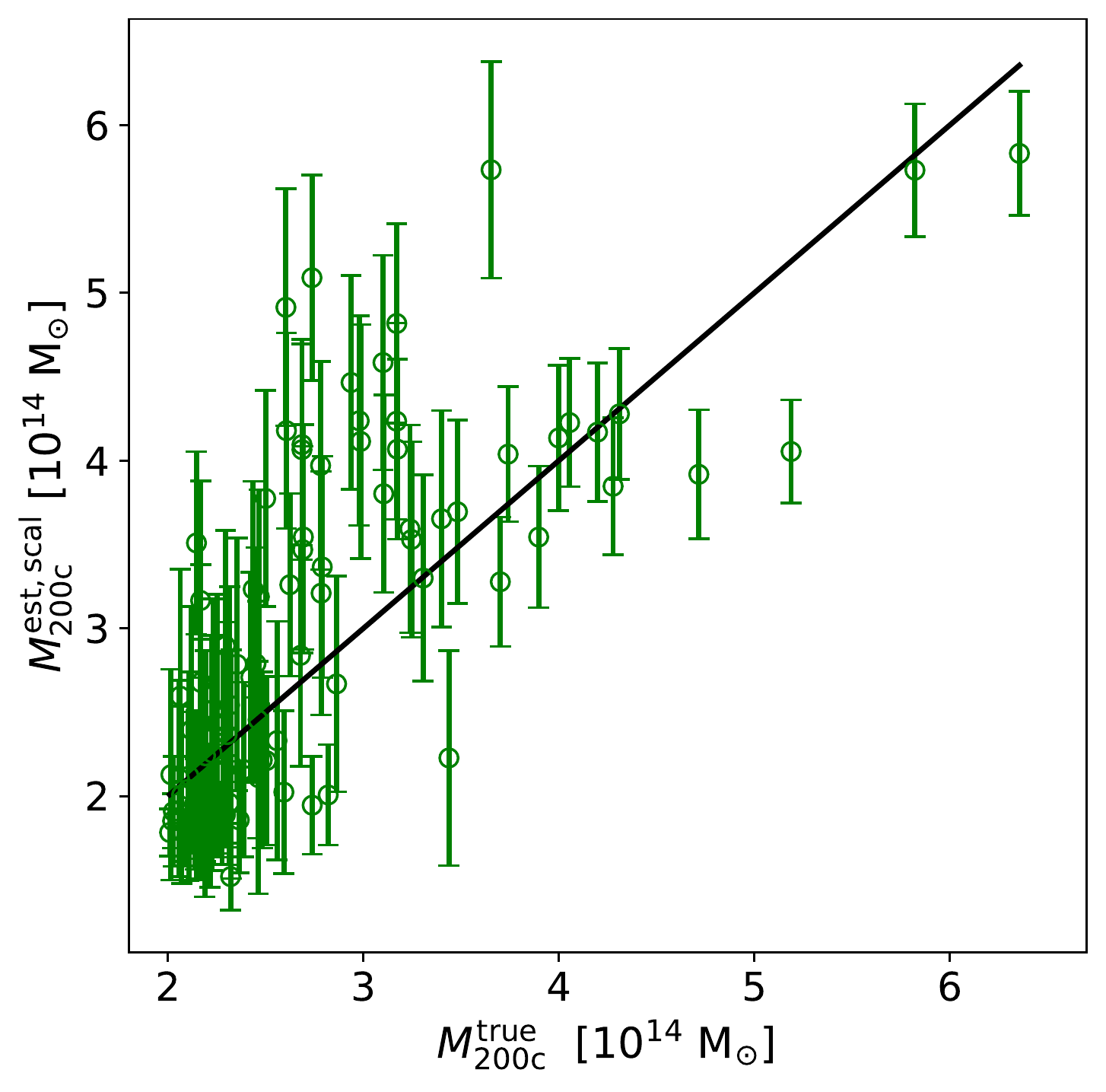}
\includegraphics[width=8.2cm, scale=0.5]{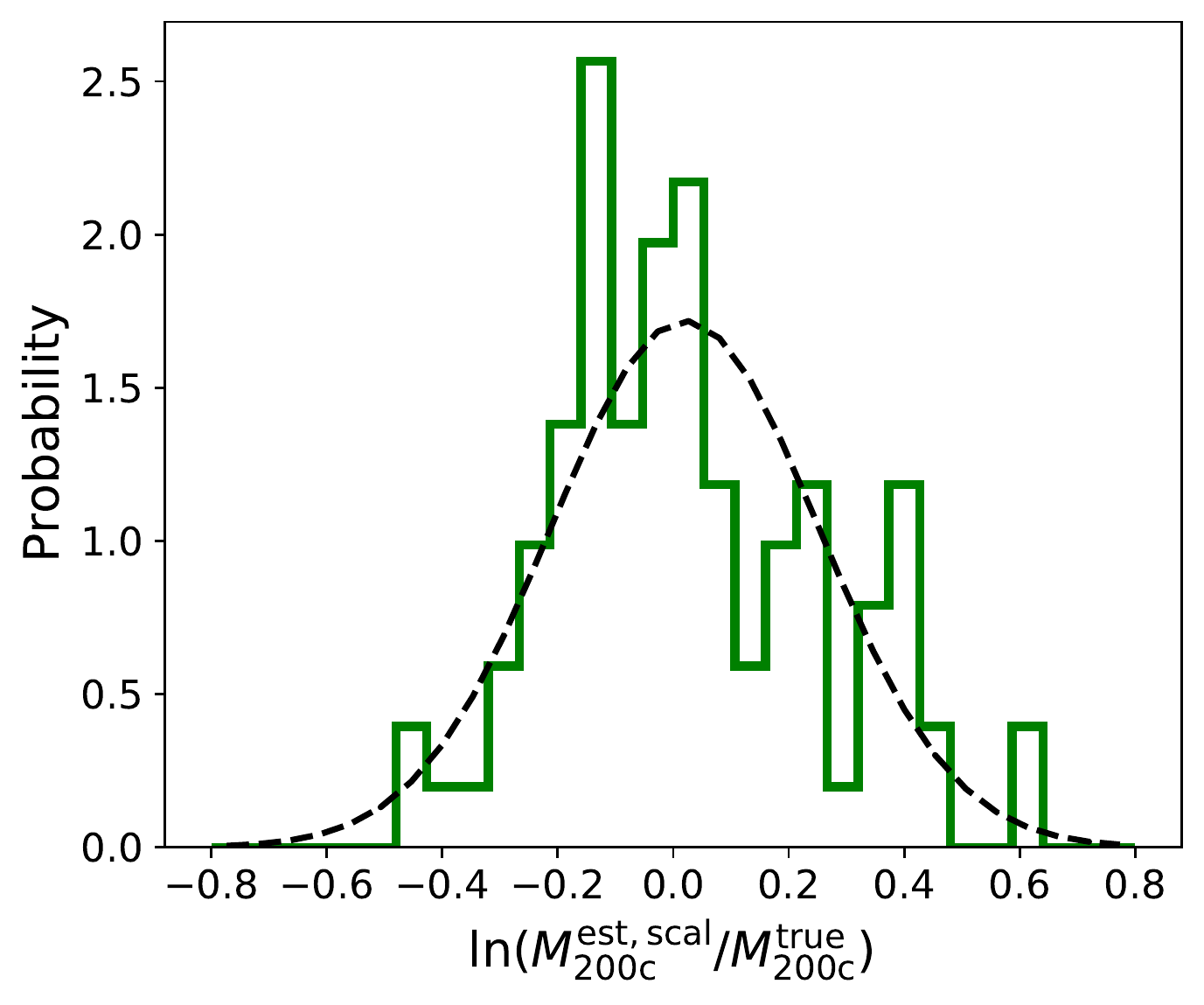}
\caption{
The deep learning model recovers cluster masses for the independent Magneticum hydrodynamical simulation. 
The top panel plots the mass estimated by the model to the true mass from the simulation for each of the 95 galaxy clusters. 
The estimated mass is scaled to account for bias due to the differences between the Arnaud and MPS scaling relations.
The black line shows the ideal where $M^{\rm est,scal}_{\rm 200c} = M^{\rm true}_{\rm 200c}$. 
The bottom panel shows the histogram of $\ln(M^{\rm est,scal}_{\rm 200c}/M^{\rm true}_{\rm 200c})$ (solid green line) for these 95 clusters. 
Fitting a Gaussian to this distribution (black dashed contours) yields a standard deviation of $\sigma = 0.232\pm0.018$, primarily due to the log-normal scatter of $\sim 0.194$ in the simulation. 
The recovered mean is $\mu=0.013\pm0.011$, consistent with no mass bias after correcting for the expected difference in scaling relations. 
This test shows that the deep learning technique can robustly recover masses from more realistic SZ signals even when trained on the simple Arnaud profile.}
\label{FIG:magneticum}
\end{figure}
\subsection{Testing Model with External Hydrodynamical Simulations}
\label{sec:magneticum}

In this section, we present our trained mResUNet model predictions for test images from the MPS, a large hydrodynamical simulation carried out as a counterpart to ongoing, multiwavelength surveys.
The details about the simulations are discussed elsewhere \citep[e.g.][]{dolag16b, gupta17b, soergel18}, and here we briefly summarize the most relevant features used in this work.

We use the two-dimensional Compton-$y$ map created by applying the so-called ‘gather approximation’ with the SPH kernel \citep{monaghan85, dolag05b}, where all gas particles that project into the target pixel contribute to the total $y$. 
The projection effects due to the uncorrelated line of sight structures are added by constructing four light cones from randomly selected slices without rotating the simulation box. 
Each light cone is a stack of 27 slices extracted from the simulation box at different redshifts.
We use these light cones to extract cutouts of 95 galaxy clusters at $z=0.67$ and $z=0.73$ with 2$\times 10^{14}~\rm M_{\odot}$ $<M_{200\rm c}<$ 7$\times 10^{14}~\rm M_{\odot}$. 
These cutouts have a resolution of $\sim 0.2^{\prime}$ per pixel and we increase it to $0.25^{\prime}$ to match with the pixel size of our training sample.
The cluster catalog for these light cones have masses defined as $M_{\rm 500c}$, that is, the mass within the region where the average mass density is 500 times the critical density of universe. 
We change this to $M_{200 \rm c}$ using a model of concentration-mass relation given by \cite{diemer15}.
We change the Compton-$y$ maps to temperature maps at 150~GHz and add them to the random realizations of CMB as well as foregrounds as described in Section~\ref{sec:SZE}.
Similar to training and validation samples, we add 5~$\rm \mu K$-arcmin white noise and convolve these maps with $1^{\prime}$ telescope beam.

Since the SZ-mass scaling relation used in training the deep learning model is different than that found in the MPS simulation \citep{gupta17b}, we should not expect the deep learning model to recover unbiased masses for the MPS simulation.
As discussed in Section~\ref{sec:systematics}, uncertainty in the SZ-mass scaling relation poses a fundamental limit to how accurately masses can be recovered from the SZ flux. 
This limit will improve as future lensing surveys improve our knowledge of the relationship. 
The interesting question to test with the MPS simulations is not whether the method is sensitive to the SZ-mass scaling relation (it is), but whether the deep learning technique can recover masses from more realistic SZ signals when trained on the simple Arnaud profile.

Thus, we rescale the estimated masses based on the scaling relation differences. 
Specifically, we scale the mass of each cluster by the factor, r:
\BE
\label{EQ:ratio}
r = \frac{F_{\rm Ar} (G_{\rm MPS} (M,z), z=0.7)}{M},
\EE
where $Y=G_{MPS}(M,z)$ is the function describing the expected $Y$ for a cluster of a given mass and redshift in the MPS simulation, and $M = F_{\rm Ar}(Y,z)$ the inverse function for the Arnaud scaling relation used in training the model. 
The redshift is fixed to $z=0.7$ as in the training set. 
Recall that the redshift in MPS is restricted to the narrow range $z \in [ 0.67, 0.73]$. 
The SZ scaling relation in the MPS is taken from Table~4 in \citep{gupta17b}. 
The reported uncertainties on the scaling relation parameters in that work are small and only lead to a small 1.7\% scatter in this factor (which we neglect). 
A caveat is that, since that work only reports the $Y^{\rm cyl}_{\rm 500c}$-$M_{\rm 500c}$ scaling relation\footnote{We note that  \citet{gupta17b} refers to `cyl' by `lc'.}, we are adjusting the $M_{200c}$ results in this work by the expected $M_{500c}$ mass ratios.
We scale the masses estimated by the deep learning model by this factor r to get re-scaled mass estimates:
\BE
\label{EQ:Mest}
M^{\rm est,scal}_{\rm 200c} = r  M^{\rm est}_{\rm 200c}.
\EE
The mean $r$ over the set of MPS clusters used is 1.287.

The top panel of Figure~\ref{FIG:magneticum} shows the scaled mass estimate plotted against the true mass of the 95 MPS galaxy clusters. 
The error bars are estimated by looking at the scatter across 100 realisations of the CMB and foregrounds that are added to the SZ signal of each cluster. 
The bottom panel of Figure~\ref{FIG:magneticum} shows the distribution of the logarithm of the ratio of the  scaled mass estimate to the true mass (solid green line). 
As in the previous section, we fit a Gaussian function to this distribution. 
We find the mean is $0.013\pm0.011$, consistent with zero, i.e. no mass bias. 
This argues that the method can accurately recover the mass from realistic SZ profiles even when the deep learning model is trained on simpler axisymmetric profiles.

In Section~\ref{sec:scatter}, we showed that the uncertainty in the recovered mass was dominated by the intrinsic scatter in the SZ-mass scaling relation. 
We now check if this is still true for the more realistic SZ profiles in the MPS simulations. 
As in Section~\ref{sec:scatter}, we would like to compare the log-normal scatter in the scaled mass estimate to the intrinsic scatter in the MPS simulation. 
For the former, the Gaussian fit to the bottom panel of Figure~\ref{FIG:magneticum} has a width $\sigma=0.232\pm0.018$. 
For the latter, \citet{gupta17b} found an intrinsic log-normal scatter of $0.159\pm 0.002$ in the $Y^{\rm cyl}_{500 \rm c}-M_{500 \rm c}$ scaling relation.
Unfortunately, that work did not look at the scaling between $Y^{\rm cyl}_{200 \rm c}$ and $M_{200 \rm c}$.
However, they did report that the scatter within $R_{200c}$ is a factor of 1.22 times larger than the scatter within $R_{\rm 500c}$ for the spherical $Y$ quantities  \citep[Table 3 in][]{gupta17b}.
Assuming that the same factor is valid for the cylindrical quantities, at $0.232\pm0.018$, the scatter in the estimated mass is only slightly larger than the intrinsic scatter of $0.194\pm 0.002$ in the simulation, with the shift marginally detected at $2.1\,\sigma$ level. 
The performance of the deep learning method appears limited by the intrinsic scatter in the SZ flux. 

\section{Conclusions}
\label{sec:conclusions}

We estimate  masses of galaxy clusters directly from simulated images of the microwave sky for the first time, using the mResUNet deep learning algorithm.
The mResUNet model is a feed-forward neural network designed for image to image regression. 
The trained mResUNet model simultaneously predicts a cluster's SZ profile and mass, directly from an image of the microwave sky at the cluster location.

We train the model using Arnaud profiles for the SZ signal added to Gaussian realisation of the CMB and astrophysical foregrounds. 
We include a 20\% log-normal scatter in the predicted SZ signal as a function of cluster mass. 
We train the model with 200 simulated images at each of eight cluster masses, with $M_{200\rm c} =$ (1, 2, 3, 4, 5, 6, 7, 8)$\times 10^{14}~\rm M_{\odot}$.
 
 We verify the trained model using different simulated images.
 We find that the trained model accurately recovers the cluster masses when the masses are within the trained range. 
 For instance, we find $M_{200 \rm c}=(1.99\pm 0.40) \times 10^{14}~\rm M_{\odot}$ for an input mass of $M_{200 \rm c}^{\rm True}=2 \times 10^{14}~\rm M_{\odot}$.
 The combined intrinsic and observational scatter is consistent with the modelled 20\% intrinsic log-normal SZ-mass scatter. 
 We test this by comparing the scatter in the recovered masses for a set of 1000 clusters with masses randomly drawn from the mass range 2$\times 10^{14}~\rm M_{\odot}$ $<M_{200\rm c}<$ 7$\times 10^{14}~\rm M_{\odot}$. 
 The fractional mass error across this set of 1000 clusters drops from $0.180\pm 0.013$ to $0.100\pm0.012$ when the log-normal SZ scatter is set to zero, proving that the SZ scatter is the main source of uncertainty. 
 
 The model does not recover the mass of clusters outside the trained mass range. 
 Unsurprisingly, for lower (higher) masses, it returns the lowest (highest) trained mass instead of the true mass. 
 
 While the model is trained on simplified SZ profiles (spherically symmetric Arnaud profiles), the trained model performs well when provided images with more realistic SZ profiles. 
 We demonstrate this by  taking $95$ galaxy cluster cutouts from the light cones of the Magneticum hydrodynamical simulation at $z=0.67$ and $z=0.73$ with 2$\times 10^{14}~\rm M_{\odot}$ $<M_{200\rm c}<$ 7$\times 10^{14}~\rm M_{\odot}$. 
 These cutouts include both more complex SZ structure from the cluster itself, as well as the added SZ contributions from other objects along nearby lines of sight. 
The model recovers the true masses of the clusters after 
correcting for the differences between the  Arnaud and MPS SZ-mass scaling relations, with a combined intrinsic and observational log-normal scatter of $0.237\pm0.018$.
 Intuitively, the model, which is trained on azimuthally symmetric SZ profiles, is analogous to taking the integrated Compton-y within a radius. 
 This test demonstrates that the deep learning method should work on actual SZ images of galaxy clusters, even if the training set does not capture the full complexity of the real SZ signal. 

In a future work, we will implement this deep learning approach to estimate the mass of galaxy clusters using the real observations of microwave sky.
Deep-learning-based mass estimation could provide an efficient way to estimate cluster masses for the sample of $>$10$^4$ galaxy clusters expected from ongoing \citep[e.g. SPT-3G, AdvancedACT][]{benson14, henderson16} and upcoming \citep[e.g. Simons Observatory, CMB-S4][]{simons19, cmbs4-19} CMB surveys. 
While requiring a much larger training and validation data sets with wider dynamic range of mass and redshift of clusters,  deep learning networks can provide  accurate mass measurements of galaxy clusters for current and future SZ surveys.

\begin{acknowledgements}

We acknowledge support from the Australian Research Council's Discovery Projects scheme (DP150103208).
We thank Raffaella Capasso, Sebastian Grandis, Brian Nord, Jo\~{a}o Caldeira, Sanjay Patil and Federico Bianchini for their helpful feedback.

\end{acknowledgements}

\bibliographystyle{aasjournal}
\bibliography{CNN_clusters}

\begin{thebibliography}{}
\expandafter\ifx\csname natexlab\endcsname\relax\def\natexlab#1{#1}\fi
\providecommand{\url}[1]{\href{#1}{#1}}
\providecommand{\dodoi}[1]{doi:~\href{http://doi.org/#1}{\nolinkurl{#1}}}
\providecommand{\doeprint}[1]{\href{http://ascl.net/#1}{\nolinkurl{http://ascl.net/#1}}}
\providecommand{\doarXiv}[1]{\href{https://arxiv.org/abs/#1}{\nolinkurl{https://arxiv.org/abs/#1}}}

\bibitem[{{Abazajian} {et~al.}(2019){Abazajian}, {Addison}, {Adshead}, {Ahmed},
  {Allen}, {Alonso}, {Alvarez}, {Anderson}, {Arnold}, {Baccigalupi}, {Bailey},
  {Barkats}, {Barron}, {Barry}, {Bartlett}, {Basu Thakur}, {Battaglia},
  {Baxter}, {Bean}, {Bebek}, {Bender}, {Benson}, {Berger}, {Bhimani},
  {Bischoff}, {Bleem}, {Bocquet}, {Boddy}, {Bonato}, {Bond}, {Borrill},
  {Bouchet}, {Brown}, {Bryan}, {Burkhart}, {Buza}, {Byrum}, {Calabrese},
  {Calafut}, {Caldwell}, {Carlstrom}, {Carron}, {Cecil}, {Challinor}, {Chang},
  {Chinone}, {Cho}, {Cooray}, {Crawford}, {Crites}, {Cukierman}, {Cyr-Racine},
  {de Haan}, {de Zotti}, {Delabrouille}, {Demarteau}, {Devlin}, {Di Valentino},
  {Dobbs}, {Duff}, {Duivenvoorden}, {Dvorkin}, {Edwards}, {Eimer}, {Errard},
  {Essinger-Hileman}, {Fabbian}, {Feng}, {Ferraro}, {Filippini}, {Flauger},
  {Flaugher}, {Fraisse}, {Frolov}, {Galitzki}, {Galli}, {Ganga}, {Gerbino},
  {Gilchriese}, {Gluscevic}, {Green}, {Grin}, {Grohs}, {Gualtieri}, {Guarino},
  {Gudmundsson}, {Habib}, {Haller}, {Halpern}, {Halverson}, {Hanany},
  {Harrington}, {Hasegawa}, {Hasselfield}, {Hazumi}, {Heitmann}, {Henderson},
  {Henning}, {Hill}, {Hlozek}, {Holder}, {Holzapfel}, {Hubmayr},
  {Huffenberger}, {Huffer}, {Hui}, {Irwin}, {Johnson}, {Johnstone}, {Jones},
  {Karkare}, {Katayama}, {Kerby}, {Kernovsky}, {Keskitalo}, {Kisner}, {Knox},
  {Kosowsky}, {Kovac}, {Kovetz}, {Kuhlmann}, {Kuo}, {Kurita}, {Kusaka},
  {Lahteenmaki}, {Lawrence}, {Lee}, {Lewis}, {Li}, {Linder}, {Loverde},
  {Lowitz}, {Madhavacheril}, {Mantz}, {Matsuda}, {Mauskopf}, {McMahon},
  {McQuinn}, {Meerburg}, {Melin}, {Meyers}, {Millea}, {Mohr}, {Moncelsi},
  {Mroczkowski}, {Mukherjee}, {M{\"u}nchmeyer}, {Nagai}, {Nagy}, {Namikawa},
  {Nati}, {Natoli}, {Negrello}, {Newburgh}, {Niemack}, {Nishino}, {Nordby},
  {Novosad}, {O'Connor}, {Obied}, {Padin}, {Pandey}, {Partridge}, {Pierpaoli},
  {Pogosian}, {Pryke}, {Puglisi}, {Racine}, {Raghunathan}, {Rahlin},
  {Rajagopalan}, {Raveri}, {Reichanadter}, {Reichardt}, {Remazeilles}, {Rocha},
  {Roe}, {Roy}, {Ruhl}, {Salatino}, {Saliwanchik}, {Schaan}, {Schillaci},
  {Schmittfull}, {Scott}, {Sehgal}, {Shandera}, {Sheehy}, {Sherwin},
  {Shirokoff}, {Simon}, {Slosar}, {Somerville}, {Spergel}, {Staggs}, {Stark},
  {Stompor}, {Story}, {Stoughton}, {Suzuki}, {Tajima}, {Teply}, {Thompson},
  {Timbie}, {Tomasi}, {Treu}, {Tristram}, {Tucker}, {Umilt{\`a}}, {van
  Engelen}, {Vieira}, {Vieregg}, {Vogelsberger}, {Wang}, {Watson}, {White},
  {Whitehorn}, {Wollack}, {Kimmy Wu}, {Xu}, {Yasini}, {Yeck}, {Yoon}, {Young},
  \& {Zonca}}]{cmbs4-19}
{Abazajian}, K., {Addison}, G., {Adshead}, P., {et~al.} 2019, arXiv e-prints,
  arXiv:1907.04473.
\newblock \doarXiv{1907.04473}

\bibitem[{{Ade} {et~al.}(2019){Ade}, {Aguirre}, {Ahmed}, {Aiola}, {Ali},
  {Alonso}, {Alvarez}, {Arnold}, {Ashton}, {Austermann}, {Awan}, {Baccigalupi},
  {Baildon}, {Barron}, {Battaglia}, {Battye}, {Baxter}, {Bazarko}, {Beall},
  {Bean}, {Beck}, {Beckman}, {Beringue}, {Bianchini}, {Boada}, {Boettger},
  {Bond}, {Borrill}, {Brown}, {Bruno}, {Bryan}, {Calabrese}, {Calafut},
  {Calisse}, {Carron}, {Challinor}, {Chesmore}, {Chinone}, {Chluba}, {Cho},
  {Choi}, {Coppi}, {Cothard}, {Coughlin}, {Crichton}, {Crowley}, {Crowley},
  {Cukierman}, {D'Ewart}, {D{\"u}nner}, {de Haan}, {Devlin}, {Dicker},
  {Didier}, {Dobbs}, {Dober}, {Duell}, {Duff}, {Duivenvoorden}, {Dunkley},
  {Dusatko}, {Errard}, {Fabbian}, {Feeney}, {Ferraro}, {Flux{\`a}}, {Freese},
  {Frisch}, {Frolov}, {Fuller}, {Fuzia}, {Galitzki}, {Gallardo}, {Tomas Galvez
  Ghersi}, {Gao}, {Gawiser}, {Gerbino}, {Gluscevic}, {Goeckner-Wald}, {Golec},
  {Gordon}, {Gralla}, {Green}, {Grigorian}, {Groh}, {Groppi}, {Guan},
  {Gudmundsson}, {Han}, {Hargrave}, {Hasegawa}, {Hasselfield}, {Hattori},
  {Haynes}, {Hazumi}, {He}, {Healy}, {Henderson}, {Hervias-Caimapo}, {Hill},
  {Hill}, {Hilton}, {Hilton}, {Hincks}, {Hinshaw}, {Hlo{\v{z}}ek}, {Ho}, {Ho},
  {Howe}, {Huang}, {Hubmayr}, {Huffenberger}, {Hughes}, {Ijjas}, {Ikape},
  {Irwin}, {Jaffe}, {Jain}, {Jeong}, {Kaneko}, {Karpel}, {Katayama}, {Keating},
  {Kernasovskiy}, {Keskitalo}, {Kisner}, {Kiuchi}, {Klein}, {Knowles},
  {Koopman}, {Kosowsky}, {Krachmalnicoff}, {Kuenstner}, {Kuo}, {Kusaka},
  {Lashner}, {Lee}, {Lee}, {Leon}, {Leung}, {Lewis}, {Li}, {Li}, {Limon},
  {Linder}, {Lopez-Caraballo}, {Louis}, {Lowry}, {Lungu}, {Madhavacheril},
  {Mak}, {Maldonado}, {Mani}, {Mates}, {Matsuda}, {Maurin}, {Mauskopf}, {May},
  {McCallum}, {McKenney}, {McMahon}, {Meerburg}, {Meyers}, {Miller},
  {Mirmelstein}, {Moodley}, {Munchmeyer}, {Munson}, {Naess}, {Nati},
  {Navaroli}, {Newburgh}, {Nguyen}, {Niemack}, {Nishino}, {Orlowski-Scherer},
  {Page}, {Partridge}, {Peloton}, {Perrotta}, {Piccirillo}, {Pisano},
  {Poletti}, {Puddu}, {Puglisi}, {Raum}, {Reichardt}, {Remazeilles},
  {Rephaeli}, {Riechers}, {Rojas}, {Roy}, {Sadeh}, {Sakurai}, {Salatino},
  {Sathyanarayana Rao}, {Schaan}, {Schmittfull}, {Sehgal}, {Seibert}, {Seljak},
  {Sherwin}, {Shimon}, {Sierra}, {Sievers}, {Sikhosana}, {Silva-Feaver},
  {Simon}, {Sinclair}, {Siritanasak}, {Smith}, {Smith}, {Spergel}, {Staggs},
  {Stein}, {Stevens}, {Stompor}, {Suzuki}, {Tajima}, {Takakura}, {Teply},
  {Thomas}, {Thorne}, {Thornton}, {Trac}, {Tsai}, {Tucker}, {Ullom},
  {Vagnozzi}, {van Engelen}, {Van Lanen}, {Van Winkle}, {Vavagiakis},
  {Verg{\`e}s}, {Vissers}, {Wagoner}, {Walker}, {Ward}, {Westbrook},
  {Whitehorn}, {Williams}, {Williams}, {Wollack}, {Xu}, {Yu}, {Yu}, {Zago},
  {Zhang}, {Zhu}, \& {Simons Observatory Collaboration}}]{simons19}
{Ade}, P., {Aguirre}, J., {Ahmed}, Z., {et~al.} 2019, \jcap, 2019, 056,
  \dodoi{10.1088/1475-7516/2019/02/056}

\bibitem[{{Alexander} {et~al.}(2019){Alexander}, {Gleyzer}, {McDonough},
  {Toomey}, \& {Usai}}]{alexander19d}
{Alexander}, S., {Gleyzer}, S., {McDonough}, E., {Toomey}, M.~W., \& {Usai}, E.
  2019, arXiv e-prints, arXiv:1909.07346.
\newblock \doarXiv{1909.07346}

\bibitem[{{Allen} {et~al.}(2019){Allen}, {Andreoni}, {Bachelet}, {Berriman},
  {Bianco}, {Biswas}, {Carrasco Kind}, {Chard}, {Cho}, {Cowperthwaite},
  {Etienne}, {George}, {Gibbs}, {Graham}, {Gropp}, {Gupta}, {Haas}, {Huerta},
  {Jennings}, {Katz}, {Khan}, {Kindratenko}, {Kramer}, {Liu}, {Mahabal},
  {McHenry}, {Miller}, {Neubauer}, {Oberlin}, {Olivas}, {Rosofsky}, {Ruiz},
  {Saxton}, {Schutz}, {Schwing}, {Seidel}, {Shapiro}, {Shen}, {Shen},
  {Sip{\H{o}}cz}, {Sun}, {Towns}, {Tsokaros}, {Wei}, {Wells}, {Williams},
  {Xiong}, \& {Zhao}}]{allen19d}
{Allen}, G., {Andreoni}, I., {Bachelet}, E., {et~al.} 2019, arXiv e-prints,
  arXiv:1902.00522.
\newblock \doarXiv{1902.00522}

\bibitem[{{Armitage} {et~al.}(2019){Armitage}, {Kay}, \& {Barnes}}]{armitage19}
{Armitage}, T.~J., {Kay}, S.~T., \& {Barnes}, D.~J. 2019, \mnras, 484, 1526,
  \dodoi{10.1093/mnras/stz039}

\bibitem[{{Arnaud} {et~al.}(2010){Arnaud}, {Pratt}, {Piffaretti},
  {B{\"o}hringer}, {Croston}, \& {Pointecouteau}}]{arnaud10}
{Arnaud}, M., {Pratt}, G.~W., {Piffaretti}, R., {et~al.} 2010, \aap, 517, A92+,
  \dodoi{10.1051/0004-6361/200913416}

\bibitem[{{Baxter} {et~al.}(2015){Baxter}, {Keisler}, {Dodelson}, {Aird},
  {Allen}, {Ashby}, {Bautz}, {Bayliss}, {Benson}, {Bleem}, {Bocquet},
  {Brodwin}, {Carlstrom}, {Chang}, {Chiu}, {Cho}, {Clocchiatti}, {Crawford},
  {Crites}, {Desai}, {Dietrich}, {de Haan}, {Dobbs}, {Foley}, {Forman},
  {George}, {Gladders}, {Gonzalez}, {Halverson}, {Harrington}, {Hennig},
  {Hoekstra}, {Holder}, {Holzapfel}, {Hou}, {Hrubes}, {Jones}, {Knox}, {Lee},
  {Leitch}, {Liu}, {Lueker}, {Luong-Van}, {Mantz}, {Marrone}, {McDonald},
  {McMahon}, {Meyer}, {Millea}, {Mocanu}, {Murray}, {Padin}, {Pryke},
  {Reichardt}, {Rest}, {Ruhl}, {Saliwanchik}, {Saro}, {Sayre}, {Schaffer},
  {Shirokoff}, {Song}, {Spieler}, {Stalder}, {Stanford}, {Staniszewski},
  {Stark}, {Story}, {van Engelen}, {Vanderlinde}, {Vieira}, {Vikhlinin},
  {Williamson}, {Zahn}, \& {Zenteno}}]{baxter15}
{Baxter}, E.~J., {Keisler}, R., {Dodelson}, S., {et~al.} 2015, \apj, 806, 247,
  \dodoi{10.1088/0004-637X/806/2/247}

\bibitem[{{Becker} \& {Kravtsov}(2011)}]{becker11}
{Becker}, M.~R., \& {Kravtsov}, A.~V. 2011, \apj, 740, 25,
  \dodoi{10.1088/0004-637X/740/1/25}

\bibitem[{{Benson} {et~al.}(2014){Benson}, {Ade}, {Ahmed}, {Allen}, {Arnold},
  {Austermann}, {Bender}, {Bleem}, {Carlstrom}, {Chang}, {Cho}, {Cliche},
  {Crawford}, {Cukierman}, {de Haan}, {Dobbs}, {Dutcher}, {Everett}, {Gilbert},
  {Halverson}, {Hanson}, {Harrington}, {Hattori}, {Henning}, {Hilton},
  {Holder}, {Holzapfel}, {Irwin}, {Keisler}, {Knox}, {Kubik}, {Kuo}, {Lee},
  {Leitch}, {Li}, {McDonald}, {Meyer}, {Montgomery}, {Myers}, {Natoli},
  {Nguyen}, {Novosad}, {Padin}, {Pan}, {Pearson}, {Reichardt}, {Ruhl},
  {Saliwanchik}, {Simard}, {Smecher}, {Sayre}, {Shirokoff}, {Stark}, {Story},
  {Suzuki}, {Thompson}, {Tucker}, {Vanderlinde}, {Vieira}, {Vikhlinin}, {Wang},
  {Yefremenko}, \& {Yoon}}]{benson14}
{Benson}, B.~A., {Ade}, P.~A.~R., {Ahmed}, Z., {et~al.} 2014, Society of
  Photo-Optical Instrumentation Engineers (SPIE) Conference Series, Vol. 9153,
  {SPT-3G: a next-generation cosmic microwave background polarization
  experiment on the South Pole telescope}, 91531P, \dodoi{10.1117/12.2057305}

\bibitem[{Biviano {et~al.}(2013)Biviano, Rosati, Balestra, Mercurio, Girardi,
  Nonino, Grillo, Scodeggio, Lemze, Kelson, \& et~al.}]{biviano13}
Biviano, A., Rosati, P., Balestra, I., {et~al.} 2013, Astronomy \&
  Astrophysics, 558, A1, \dodoi{10.1051/0004-6361/201321955}

\bibitem[{{Bleem} {et~al.}(2019){Bleem}, {Bocquet}, {Stalder}, {Gladders},
  {Ade}, {Allen}, {Anderson}, {Annis}, {Ashby}, {Austermann}, {Avila}, {Avva},
  {Bayliss}, {Beall}, {Bechtol}, {Bender}, {Benson}, {Bertin}, {Bianchini},
  {Blake}, {Brodwin}, {Brooks}, {Buckley-Geer}, {Burke}, {Carlstrom}, {Carnero
  Rosell}, {Carrasco Kind}, {Carretero}, {Chang}, {Chiang}, {Citron}, {Corbett
  Moran}, {Costanzi}, {Crawford}, {Crites}, {da Costa}, {de Haan}, {De
  Vicente}, {Desai}, {Diehl}, {Dietrich}, {Dobbs}, {Eifler}, {Everett},
  {Flaugher}, {Floyd}, {Frieman}, {Gallicchio}, {Garc{\'\i}a-Bellido},
  {George}, {Gerdes}, {Gilbert}, {Gruen}, {Gruendl}, {Gschwend}, {Gupta},
  {Gutierrez}, {Halverson}, {Harrington}, {Henning}, {Heymans}, {Holder},
  {Hollowood}, {Holzapfel}, {Honscheid}, {Hrubes}, {Huang}, {Hubmayr}, {Irwin},
  {James}, {Jeltema}, {Joudaki}, {Khullar}, {Klein}, {Knox}, {Kuropatkin},
  {Lee}, {Li}, {Lidman}, {Lowitz}, {MacCrann}, {Mahler}, {Maia}, {Marshall},
  {McDonald}, {McMahon}, {Melchior}, {Menanteau}, {Meyer}, {Miquel}, {Mocanu},
  {Mohr}, {Montgomery}, {Nadolski}, {Natoli}, {Nibarger}, {Noble}, {Novosad},
  {Padin}, {Palmese}, {Parkinson}, {Patil}, {Paz-Chinch{\'o}n}, {Plazas},
  {Pryke}, {Ramachandra}, {Reichardt}, {Remolina Gonz{\'a}lez}, {Romer},
  {Roodman}, {Ruhl}, {Rykoff}, {Saliwanchik}, {Sanchez}, {Saro}, {Sayre},
  {Schaffer}, {Schrabback}, {Serrano}, {Sharon}, {Sievers}, {Smecher}, {Smith},
  {Soares-Santos}, {Stark}, {Story}, {Suchyta}, {Tarle}, {Tucker},
  {Vanderlinde}, {Veach}, {Vieira}, {Wang}, {Weller}, {Whitehorn}, {Wu},
  {Yefremenko}, \& {Zhang}}]{bleem19}
{Bleem}, L.~E., {Bocquet}, S., {Stalder}, B., {et~al.} 2019, arXiv e-prints,
  arXiv:1910.04121.
\newblock \doarXiv{1910.04121}

\bibitem[{{Bocquet} {et~al.}(2015){Bocquet}, {Saro}, {Mohr}, {Aird}, {Ashby},
  {Bautz}, {Bayliss}, {Bazin}, {Benson}, {Bleem}, {Brodwin}, {Carlstrom},
  {Chang}, {Chiu}, {Cho}, {Clocchiatti}, {Crawford}, {Crites}, {Desai}, {de
  Haan}, {Dietrich}, {Dobbs}, {Foley}, {Forman}, {Gangkofner}, {George},
  {Gladders}, {Gonzalez}, {Halverson}, {Hennig}, {Hlavacek-Larrondo}, {Holder},
  {Holzapfel}, {Hrubes}, {Jones}, {Keisler}, {Knox}, {Lee}, {Leitch}, {Liu},
  {Lueker}, {Luong-Van}, {Marrone}, {McDonald}, {McMahon}, {Meyer}, {Mocanu},
  {Murray}, {Padin}, {Pryke}, {Reichardt}, {Rest}, {Ruel}, {Ruhl},
  {Saliwanchik}, {Sayre}, {Schaffer}, {Shirokoff}, {Spieler}, {Stalder},
  {Stanford}, {Staniszewski}, {Stark}, {Story}, {Stubbs}, {Vanderlinde},
  {Vieira}, {Vikhlinin}, {Williamson}, {Zahn}, \& {Zenteno}}]{bocquet15}
{Bocquet}, S., {Saro}, A., {Mohr}, J.~J., {et~al.} 2015, \apj, 799, 214,
  \dodoi{10.1088/0004-637X/799/2/214}

\bibitem[{{Bocquet} {et~al.}(2019){Bocquet}, {Dietrich}, {Schrabback}, {Bleem},
  {Klein}, {Allen}, {Applegate}, {Ashby}, {Bautz}, {Bayliss}, {Benson},
  {Brodwin}, {Bulbul}, {Canning}, {Capasso}, {Carlstrom}, {Chang}, {Chiu},
  {Cho}, {Clocchiatti}, {Crawford}, {Crites}, {de Haan}, {Desai}, {Dobbs},
  {Foley}, {Forman}, {Garmire}, {George}, {Gladders}, {Gonzalez}, {Grandis},
  {Gupta}, {Halverson}, {Hlavacek-Larrondo}, {Hoekstra}, {Holder}, {Holzapfel},
  {Hou}, {Hrubes}, {Huang}, {Jones}, {Khullar}, {Knox}, {Kraft}, {Lee}, {von
  der Linden}, {Luong-Van}, {Mantz}, {Marrone}, {McDonald}, {McMahon}, {Meyer},
  {Mocanu}, {Mohr}, {Morris}, {Padin}, {Patil}, {Pryke}, {Rapetti},
  {Reichardt}, {Rest}, {Ruhl}, {Saliwanchik}, {Saro}, {Sayre}, {Schaffer},
  {Shirokoff}, {Stalder}, {Stanford}, {Staniszewski}, {Stark}, {Story},
  {Strazzullo}, {Stubbs}, {Vanderlinde}, {Vieira}, {Vikhlinin}, {Williamson},
  \& {Zenteno}}]{bocquet19}
{Bocquet}, S., {Dietrich}, J.~P., {Schrabback}, T., {et~al.} 2019, \apj, 878,
  55, \dodoi{10.3847/1538-4357/ab1f10}

\bibitem[{{Bottrell} {et~al.}(2019){Bottrell}, {Hani}, {Teimoorinia},
  {Ellison}, {Moreno}, {Torrey}, {Hayward}, {Thorp}, {Simard}, \&
  {Hernquist}}]{bottrell19d}
{Bottrell}, C., {Hani}, M.~H., {Teimoorinia}, H., {et~al.} 2019, \mnras, 490,
  5390, \dodoi{10.1093/mnras/stz2934}

\bibitem[{{Caldeira} {et~al.}(2019){Caldeira}, {Wu}, {Nord}, {Avestruz},
  {Trivedi}, \& {Story}}]{caldeira19d}
{Caldeira}, J., {Wu}, W.~L.~K., {Nord}, B., {et~al.} 2019, Astronomy and
  Computing, 28, 100307, \dodoi{10.1016/j.ascom.2019.100307}

\bibitem[{{Capasso} {et~al.}(2019){Capasso}, {Saro}, {Mohr}, {Biviano},
  {Bocquet}, {Strazzullo}, {Grandis}, {Applegate}, {Bayliss}, {Benson},
  {Bleem}, {Brodwin}, {Bulbul}, {Carlstrom}, {Chiu}, {Dietrich}, {Gupta}, {de
  Haan}, {Hlavacek-Larrondo}, {Klein}, {von der Linden}, {McDonald}, {Rapetti},
  {Reichardt}, {Sharon}, {Stalder}, {Stanford}, {Stark}, {Stern}, \&
  {Zenteno}}]{capasso19}
{Capasso}, R., {Saro}, A., {Mohr}, J.~J., {et~al.} 2019, \mnras, 482, 1043,
  \dodoi{10.1093/mnras/sty2645}

\bibitem[{{Carlstrom} {et~al.}(2011){Carlstrom}, {Ade}, {Aird}, {Benson},
  {Bleem}, {Busetti}, {Chang}, {Chauvin}, {Cho}, {Crawford}, {Crites}, {Dobbs},
  {Halverson}, {Heimsath}, {Holzapfel}, {Hrubes}, {Joy}, {Keisler}, {Lanting},
  {Lee}, {Leitch}, {Leong}, {Lu}, {Lueker}, {Luong-van}, {McMahon}, {Mehl},
  {Meyer}, {Mohr}, {Montroy}, {Padin}, {Plagge}, {Pryke}, {Ruhl}, {Schaffer},
  {Schwan}, {Shirokoff}, {Spieler}, {Staniszewski}, {Stark}, {Tucker},
  {Vanderlinde}, {Vieira}, \& {Williamson}}]{carlstrom11}
{Carlstrom}, J.~E., {Ade}, P.~A.~R., {Aird}, K.~A., {et~al.} 2011, \pasp, 123,
  568, \dodoi{10.1086/659879}

\bibitem[{{Chen} {et~al.}(2016){Chen}, {Papandreou}, {Kokkinos}, {Murphy}, \&
  {Yuille}}]{chen16d}
{Chen}, L.-C., {Papandreou}, G., {Kokkinos}, I., {Murphy}, K., \& {Yuille},
  A.~L. 2016, arXiv e-prints, arXiv:1606.00915.
\newblock \doarXiv{1606.00915}

\bibitem[{{Chen} {et~al.}(2017){Chen}, {Papandreou}, {Schroff}, \&
  {Adam}}]{chen17d}
{Chen}, L.-C., {Papandreou}, G., {Schroff}, F., \& {Adam}, H. 2017, arXiv
  e-prints, arXiv:1706.05587.
\newblock \doarXiv{1706.05587}

\bibitem[{{Corless} \& {King}(2009)}]{corless09}
{Corless}, V.~L., \& {King}, L.~J. 2009, \mnras, 396, 315,
  \dodoi{10.1111/j.1365-2966.2009.14542.x}

\bibitem[{de~Haan {et~al.}(2016)de~Haan, Benson, Bleem, Allen, Applegate,
  Ashby, Bautz, Bayliss, Bocquet, Brodwin, Carlstrom, Chang, Chiu, Cho,
  Clocchiatti, Crawford, Crites, Desai, Dietrich, Dobbs, Doucouliagos, Foley,
  Forman, Garmire, George, Gladders, Gonzalez, Gupta, Halverson,
  Hlavacek-Larrondo, Hoekstra, Holder, Holzapfel, Hou, Hrubes, Huang, Jones,
  Keisler, Knox, Lee, Leitch, von~der Linden, Luong-Van, Mantz, Marrone,
  McDonald, McMahon, Meyer, Mocanu, Mohr, Murray, Padin, Pryke, Rapetti,
  Reichardt, Rest, Ruel, Ruhl, Saliwanchik, Saro, Sayre, Schaffer, Schrabback,
  Shirokoff, Song, Spieler, Stalder, Stanford, Staniszewski, Stark, Story,
  Stubbs, Vanderlinde, Vieira, Vikhlinin, Williamson, \& Zenteno}]{dehaan16}
de~Haan, T., Benson, B.~A., Bleem, L.~E., {et~al.} 2016, The Astrophysical
  Journal, 832, 95.
\newblock \url{http://stacks.iop.org/0004-637X/832/i=1/a=95}

\bibitem[{{Diemer} \& {Kravtsov}(2015)}]{diemer15}
{Diemer}, B., \& {Kravtsov}, A.~V. 2015, \apj, 799, 108,
  \dodoi{10.1088/0004-637X/799/1/108}

\bibitem[{{Dietrich} {et~al.}(2019){Dietrich}, {Bocquet}, {Schrabback},
  {Applegate}, {Hoekstra}, {Grandis}, {Mohr}, {Allen}, {Bayliss}, {Benson},
  {Bleem}, {Brodwin}, {Bulbul}, {Capasso}, {Chiu}, {Crawford}, {Gonzalez}, {de
  Haan}, {Klein}, {von der Linden}, {Mantz}, {Marrone}, {McDonald},
  {Raghunathan}, {Rapetti}, {Reichardt}, {Saro}, {Stalder}, {Stark}, {Stern},
  \& {Stubbs}}]{dietrich19}
{Dietrich}, J.~P., {Bocquet}, S., {Schrabback}, T., {et~al.} 2019, \mnras, 483,
  2871, \dodoi{10.1093/mnras/sty3088}

\bibitem[{{Dolag} {et~al.}(2016){Dolag}, {Komatsu}, \& {Sunyaev}}]{dolag16b}
{Dolag}, K., {Komatsu}, E., \& {Sunyaev}, R. 2016, \mnras, 463, 1797,
  \dodoi{10.1093/mnras/stw2035}

\bibitem[{{Dolag} {et~al.}(2005){Dolag}, {Vazza}, {Brunetti}, \&
  {Tormen}}]{dolag05b}
{Dolag}, K., {Vazza}, F., {Brunetti}, G., \& {Tormen}, G. 2005, \mnras, 364,
  753, \dodoi{10.1111/j.1365-2966.2005.09630.x}

\bibitem[{{Drozdzal} {et~al.}(2016){Drozdzal}, {Vorontsov}, {Chartrand},
  {Kadoury}, \& {Pal}}]{drozdzal16d}
{Drozdzal}, M., {Vorontsov}, E., {Chartrand}, G., {Kadoury}, S., \& {Pal}, C.
  2016, arXiv e-prints, arXiv:1608.04117.
\newblock \doarXiv{1608.04117}

\bibitem[{{Dumoulin} \& {Visin}(2016)}]{dumoulin16d}
{Dumoulin}, V., \& {Visin}, F. 2016, arXiv e-prints, arXiv:1603.07285.
\newblock \doarXiv{1603.07285}

\bibitem[{{Fluri} {et~al.}(2019){Fluri}, {Kacprzak}, {Lucchi}, {Refregier},
  {Amara}, {Hofmann}, \& {Schneider}}]{fluri19d}
{Fluri}, J., {Kacprzak}, T., {Lucchi}, A., {et~al.} 2019, \prd, 100, 063514,
  \dodoi{10.1103/PhysRevD.100.063514}

\bibitem[{{Fowler} {et~al.}(2007){Fowler}, {Niemack}, {Dicker}, {Aboobaker},
  {Ade}, {Battistelli}, {Devlin}, {Fisher}, {Halpern}, {Hargrave}, {Hincks},
  {Kaul}, {Klein}, {Lau}, {Limon}, {Marriage}, {Mauskopf}, {Page}, {Staggs},
  {Swetz}, {Switzer}, {Thornton}, \& {Tucker}}]{fowler07}
{Fowler}, J.~W., {Niemack}, M.~D., {Dicker}, S.~R., {et~al.} 2007, \ao, 46,
  3444, \dodoi{10.1364/AO.46.003444}

\bibitem[{{George} \& {Huerta}(2018)}]{george18d}
{George}, D., \& {Huerta}, E.~A. 2018, \prd, 97, 044039,
  \dodoi{10.1103/PhysRevD.97.044039}

\bibitem[{{George} {et~al.}(2015){George}, {Reichardt}, {Aird}, {Benson},
  {Bleem}, {Carlstrom}, {Chang}, {Cho}, {Crawford}, {Crites}, {de Haan},
  {Dobbs}, {Dudley}, {Halverson}, {Harrington}, {Holder}, {Holzapfel}, {Hou},
  {Hrubes}, {Keisler}, {Knox}, {Lee}, {Leitch}, {Lueker}, {Luong-Van},
  {McMahon}, {Mehl}, {Meyer}, {Millea}, {Mocanu}, {Mohr}, {Montroy}, {Padin},
  {Plagge}, {Pryke}, {Ruhl}, {Schaffer}, {Shaw}, {Shirokoff}, {Spieler},
  {Staniszewski}, {Stark}, {Story}, {van Engelen}, {Vanderlinde}, {Vieira},
  {Williamson}, \& {Zahn}}]{george15}
{George}, E.~M., {Reichardt}, C.~L., {Aird}, K.~A., {et~al.} 2015, \apj, 799,
  177, \dodoi{10.1088/0004-637X/799/2/177}

\bibitem[{{Green} {et~al.}(2019){Green}, {Ntampaka}, {Nagai}, {Lovisari},
  {Dolag}, {Eckert}, \& {ZuHone}}]{green19}
{Green}, S.~B., {Ntampaka}, M., {Nagai}, D., {et~al.} 2019, \apj, 884, 33,
  \dodoi{10.3847/1538-4357/ab426f}

\bibitem[{{Gruen} {et~al.}(2014){Gruen}, {Seitz}, {Brimioulle}, {Kosyra},
  {Koppenhoefer}, {Lee}, {Bender}, {Riffeser}, {Eichner}, {Weidinger}, \&
  {Bierschenk}}]{gruen14}
{Gruen}, D., {Seitz}, S., {Brimioulle}, F., {et~al.} 2014, \mnras, 442, 1507,
  \dodoi{10.1093/mnras/stu949}

\bibitem[{{Gu} {et~al.}(2015){Gu}, {Wang}, {Kuen}, {Ma}, {Shahroudy}, {Shuai},
  {Liu}, {Wang}, {Wang}, {Wang}, {Cai}, \& {Chen}}]{gu15d}
{Gu}, J., {Wang}, Z., {Kuen}, J., {et~al.} 2015, arXiv e-prints,
  arXiv:1512.07108.
\newblock \doarXiv{1512.07108}

\bibitem[{{Gupta} {et~al.}(2017){Gupta}, {Saro}, {Mohr}, {Dolag}, \&
  {Liu}}]{gupta17b}
{Gupta}, N., {Saro}, A., {Mohr}, J.~J., {Dolag}, K., \& {Liu}, J. 2017, \mnras,
  469, 3069, \dodoi{10.1093/mnras/stx715}

\bibitem[{{Hasselfield} {et~al.}(2013){Hasselfield}, {Hilton}, {Marriage},
  {Addison}, {Barrientos}, {Battaglia}, {Battistelli}, {Bond}, {Crichton},
  {Das}, {Devlin}, {Dicker}, {Dunkley}, {Dunner}, {Fowler}, {Gralla}, {Hajian},
  {Halpern}, {Hincks}, {Hlozek}, {Hughes}, {Infante}, {Irwin}, {Kosowsky},
  {Marsden}, {Menanteau}, {Moodley}, {Niemack}, {Nolta}, {Page}, {Partridge},
  {Reese}, {Schmitt}, {Sehgal}, {Sherwin}, {Sievers}, {Sif{\'o}n}, {Spergel},
  {Staggs}, {Swetz}, {Switzer}, {Thornton}, {Trac}, \&
  {Wollack}}]{hasselfield13}
{Hasselfield}, M., {Hilton}, M., {Marriage}, T.~A., {et~al.} 2013, ArXiv
  e-prints.
\newblock \doarXiv{1301.0816}

\bibitem[{{He} {et~al.}(2015){He}, {Zhang}, {Ren}, \& {Sun}}]{he15d}
{He}, K., {Zhang}, X., {Ren}, S., \& {Sun}, J. 2015, arXiv e-prints,
  arXiv:1512.03385.
\newblock \doarXiv{1512.03385}

\bibitem[{{Henderson} {et~al.}(2016){Henderson}, {Allison}, {Austermann},
  {Baildon}, {Battaglia}, {Beall}, {Becker}, {De Bernardis}, {Bond},
  {Calabrese}, {Choi}, {Coughlin}, {Crowley}, {Datta}, {Devlin}, {Duff},
  {Dunkley}, {D{\"u}nner}, {van Engelen}, {Gallardo}, {Grace}, {Hasselfield},
  {Hills}, {Hilton}, {Hincks}, {Hlo{\^z}ek}, {Ho}, {Hubmayr}, {Huffenberger},
  {Hughes}, {Irwin}, {Koopman}, {Kosowsky}, {Li}, {McMahon}, {Munson}, {Nati},
  {Newburgh}, {Niemack}, {Niraula}, {Page}, {Pappas}, {Salatino}, {Schillaci},
  {Schmitt}, {Sehgal}, {Sherwin}, {Sievers}, {Simon}, {Spergel}, {Staggs},
  {Stevens}, {Thornton}, {Van Lanen}, {Vavagiakis}, {Ward}, \&
  {Wollack}}]{henderson16}
{Henderson}, S.~W., {Allison}, R., {Austermann}, J., {et~al.} 2016, Journal of
  Low Temperature Physics, 184, 772, \dodoi{10.1007/s10909-016-1575-z}

\bibitem[{{Henning} {et~al.}(2018){Henning}, {Sayre}, {Reichardt}, {Ade},
  {Anderson}, {Austermann}, {Beall}, {Bender}, {Benson}, {Bleem}, {Carlstrom},
  {Chang}, {Chiang}, {Cho}, {Citron}, {Corbett Moran}, {Crawford}, {Crites},
  {de Haan}, {Dobbs}, {Everett}, {Gallicchio}, {George}, {Gilbert},
  {Halverson}, {Harrington}, {Hilton}, {Holder}, {Holzapfel}, {Hoover}, {Hou},
  {Hrubes}, {Huang}, {Hubmayr}, {Irwin}, {Keisler}, {Knox}, {Lee}, {Leitch},
  {Li}, {Lowitz}, {Manzotti}, {McMahon}, {Meyer}, {Mocanu}, {Montgomery},
  {Nadolski}, {Natoli}, {Nibarger}, {Novosad}, {Padin}, {Pryke}, {Ruhl},
  {Saliwanchik}, {Schaffer}, {Sievers}, {Smecher}, {Stark}, {Story}, {Tucker},
  {Vanderlinde}, {Veach}, {Vieira}, {Wang}, {Whitehorn}, {Wu}, \&
  {Yefremenko}}]{henning18}
{Henning}, J.~W., {Sayre}, J.~T., {Reichardt}, C.~L., {et~al.} 2018, \apj, 852,
  97, \dodoi{10.3847/1538-4357/aa9ff4}

\bibitem[{{Hilton} {et~al.}(2018){Hilton}, {Hasselfield}, {Sif{\'o}n},
  {Battaglia}, {Aiola}, {Bharadwaj}, {Bond}, {Choi}, {Crichton}, {Datta},
  {Devlin}, {Dunkley}, {D{\"u}nner}, {Gallardo}, {Gralla}, {Hincks}, {Ho},
  {Hubmayr}, {Huffenberger}, {Hughes}, {Koopman}, {Kosowsky}, {Louis},
  {Madhavacheril}, {Marriage}, {Maurin}, {McMahon}, {Miyatake}, {Moodley},
  {N{\ae}ss}, {Nati}, {Newburgh}, {Niemack}, {Oguri}, {Page}, {Partridge},
  {Schmitt}, {Sievers}, {Spergel}, {Staggs}, {Trac}, {van Engelen},
  {Vavagiakis}, \& {Wollack}}]{hilton18}
{Hilton}, M., {Hasselfield}, M., {Sif{\'o}n}, C., {et~al.} 2018, \apjs, 235,
  20, \dodoi{10.3847/1538-4365/aaa6cb}

\bibitem[{{Hinton} {et~al.}(2012){Hinton}, {Srivastava}, {Krizhevsky},
  {Sutskever}, \& {Salakhutdinov}}]{hinton12d}
{Hinton}, G.~E., {Srivastava}, N., {Krizhevsky}, A., {Sutskever}, I., \&
  {Salakhutdinov}, R.~R. 2012, arXiv e-prints, arXiv:1207.0580.
\newblock \doarXiv{1207.0580}

\bibitem[{{Ho} {et~al.}(2019){Ho}, {Rau}, {Ntampaka}, {Farahi}, {Trac}, \&
  {P{\'o}czos}}]{ho19}
{Ho}, M., {Rau}, M.~M., {Ntampaka}, M., {et~al.} 2019, \apj, 887, 25,
  \dodoi{10.3847/1538-4357/ab4f82}

\bibitem[{{Hoekstra} {et~al.}(2015){Hoekstra}, {Herbonnet}, {Muzzin}, {Babul},
  {Mahdavi}, {Viola}, \& {Cacciato}}]{hoekstra15}
{Hoekstra}, H., {Herbonnet}, R., {Muzzin}, A., {et~al.} 2015, \mnras, 449, 685,
  \dodoi{10.1093/mnras/stv275}

\bibitem[{{Huang} {et~al.}(2019){Huang}, {Bleem}, {Stalder}, {Ade}, {Allen},
  {Anderson}, {Austermann}, {Avva}, {Beall}, {Bender}, {Benson}, {Bianchini},
  {Bocquet}, {Brodwin}, {Carlstrom}, {Chang}, {Chiang}, {Citron}, {Corbett
  Moran}, {Crawford}, {Crites}, {de Haan}, {Dobbs}, {Everett}, {Floyd},
  {Gallicchio}, {George}, {Gilbert}, {Gladders}, {Guns}, {Gupta}, {Halverson},
  {Harrington}, {Henning}, {Hilton}, {Holder}, {Holzapfel}, {Hrubes},
  {Hubmayr}, {Irwin}, {Khullar}, {Knox}, {Lee}, {Li}, {Lowitz}, {McDonald},
  {McMahon}, {Meyer}, {Mocanu}, {Montgomery}, {Nadolski}, {Natoli}, {Nibarger},
  {Noble}, {Novosad}, {Padin}, {Patil}, {Pryke}, {Reichardt}, {Ruhl},
  {Saliwanchik}, {Saro}, {Sayre}, {Schaffer}, {Sharon}, {Sievers}, {Smecher},
  {Stark}, {Story}, {Tucker}, {Vanderlinde}, {Veach}, {Vieira}, {Wang},
  {Whitehorn}, {Wu}, \& {Yefremenko}}]{huang19}
{Huang}, N., {Bleem}, L.~E., {Stalder}, B., {et~al.} 2019, arXiv e-prints,
  arXiv:1907.09621.
\newblock \doarXiv{1907.09621}

\bibitem[{{Ioffe} \& {Szegedy}(2015)}]{Ioffe15d}
{Ioffe}, S., \& {Szegedy}, C. 2015, arXiv e-prints, arXiv:1502.03167.
\newblock \doarXiv{1502.03167}

\bibitem[{{Johnston} {et~al.}(2007){Johnston}, {Sheldon}, {Wechsler}, {Rozo},
  {Koester}, {Frieman}, {McKay}, {Evrard}, {Becker}, \& {Annis}}]{johnston07}
{Johnston}, D.~E., {Sheldon}, E.~S., {Wechsler}, R.~H., {et~al.} 2007, ArXiv
  e-prints.
\newblock \doarXiv{0709.1159}

\bibitem[{{Kayalibay} {et~al.}(2017){Kayalibay}, {Jensen}, \& {van der
  Smagt}}]{kayalibay17}
{Kayalibay}, B., {Jensen}, G., \& {van der Smagt}, P. 2017, arXiv e-prints,
  arXiv:1701.03056.
\newblock \doarXiv{1701.03056}

\bibitem[{{Kingma} \& {Ba}(2014)}]{kingma14d}
{Kingma}, D.~P., \& {Ba}, J. 2014, arXiv e-prints, arXiv:1412.6980.
\newblock \doarXiv{1412.6980}

\bibitem[{{Klambauer} {et~al.}(2017){Klambauer}, {Unterthiner}, {Mayr}, \&
  {Hochreiter}}]{klambauer17d}
{Klambauer}, G., {Unterthiner}, T., {Mayr}, A., \& {Hochreiter}, S. 2017, arXiv
  e-prints, arXiv:1706.02515.
\newblock \doarXiv{1706.02515}

\bibitem[{{Laureijs} {et~al.}(2011){Laureijs}, {Amiaux}, {Arduini},
  {Augu{\`e}res}, {Brinchmann}, {Cole}, {Cropper}, {Dabin}, {Duvet}, {Ealet},
  {Garilli}, {Gondoin}, {Guzzo}, {Hoar}, {Hoekstra}, {Holmes}, {Kitching},
  {Maciaszek}, {Mellier}, {Pasian}, {Percival}, {Rhodes}, {Saavedra Criado},
  {Sauvage}, {Scaramella}, {Valenziano}, {Warren}, {Bender}, {Castander},
  {Cimatti}, {Le F{\`e}vre}, {Kurki-Suonio}, {Levi}, {Lilje}, {Meylan},
  {Nichol}, {Pedersen}, {Popa}, {Rebolo Lopez}, {Rix}, {Rottgering},
  {Zeilinger}, {Grupp}, {Hudelot}, {Massey}, {Meneghetti}, {Miller}, {Paltani},
  {Paulin-Henriksson}, {Pires}, {Saxton}, {Schrabback}, {Seidel}, {Walsh},
  {Aghanim}, {Amendola}, {Bartlett}, {Baccigalupi}, {Beaulieu}, {Benabed},
  {Cuby}, {Elbaz}, {Fosalba}, {Gavazzi}, {Helmi}, {Hook}, {Irwin}, {Kneib},
  {Kunz}, {Mannucci}, {Moscardini}, {Tao}, {Teyssier}, {Weller}, {Zamorani},
  {Zapatero Osorio}, {Boulade}, {Foumond}, {Di Giorgio}, {Guttridge}, {James},
  {Kemp}, {Martignac}, {Spencer}, {Walton}, {Bl{\"u}mchen}, {Bonoli},
  {Bortoletto}, {Cerna}, {Corcione}, {Fabron}, {Jahnke}, {Ligori}, {Madrid},
  {Martin}, {Morgante}, {Pamplona}, {Prieto}, {Riva}, {Toledo}, {Trifoglio},
  {Zerbi}, {Abdalla}, {Douspis}, {Grenet}, {Borgani}, {Bouwens}, {Courbin},
  {Delouis}, {Dubath}, {Fontana}, {Frailis}, {Grazian}, {Koppenh{\"o}fer},
  {Mansutti}, {Melchior}, {Mignoli}, {Mohr}, {Neissner}, {Noddle}, {Poncet},
  {Scodeggio}, {Serrano}, {Shane}, {Starck}, {Surace}, {Taylor},
  {Verdoes-Kleijn}, {Vuerli}, {Williams}, {Zacchei}, {Altieri}, {Escudero
  Sanz}, {Kohley}, {Oosterbroek}, {Astier}, {Bacon}, {Bardelli}, {Baugh},
  {Bellagamba}, {Benoist}, {Bianchi}, {Biviano}, {Branchini}, {Carbone},
  {Cardone}, {Clements}, {Colombi}, {Conselice}, {Cresci}, {Deacon}, {Dunlop},
  {Fedeli}, {Fontanot}, {Franzetti}, {Giocoli}, {Garcia-Bellido}, {Gow},
  {Heavens}, {Hewett}, {Heymans}, {Holland}, {Huang}, {Ilbert}, {Joachimi},
  {Jennins}, {Kerins}, {Kiessling}, {Kirk}, {Kotak}, {Krause}, {Lahav}, {van
  Leeuwen}, {Lesgourgues}, {Lombardi}, {Magliocchetti}, {Maguire}, {Majerotto},
  {Maoli}, {Marulli}, {Maurogordato}, {McCracken}, {McLure}, {Melchiorri},
  {Merson}, {Moresco}, {Nonino}, {Norberg}, {Peacock}, {Pello}, {Penny},
  {Pettorino}, {Di Porto}, {Pozzetti}, {Quercellini}, {Radovich}, {Rassat},
  {Roche}, {Ronayette}, {Rossetti}, {Sartoris}, {Schneider}, {Semboloni},
  {Serjeant}, {Simpson}, {Skordis}, {Smadja}, {Smartt}, {Spano}, {Spiro},
  {Sullivan}, {Tilquin}, {Trotta}, {Verde}, {Wang}, {Williger}, {Zhao},
  {Zoubian}, \& {Zucca}}]{laurejis11}
{Laureijs}, R., {Amiaux}, J., {Arduini}, S., {et~al.} 2011, arXiv e-prints,
  arXiv:1110.3193.
\newblock \doarXiv{1110.3193}

\bibitem[{{Le Brun} {et~al.}(2017){Le Brun}, {McCarthy}, {Schaye}, \&
  {Ponman}}]{lebrun17}
{Le Brun}, A. M.~C., {McCarthy}, I.~G., {Schaye}, J., \& {Ponman}, T.~J. 2017,
  \mnras, 466, 4442, \dodoi{10.1093/mnras/stw3361}

\bibitem[{{Lewis} {et~al.}(2000){Lewis}, {Challinor}, \& {Lasenby}}]{lewis00}
{Lewis}, A., {Challinor}, A., \& {Lasenby}, A. 2000, \apj, 538, 473,
  \dodoi{10.1086/309179}

\bibitem[{{LSST Science Collaboration} {et~al.}(2009){LSST Science
  Collaboration}, {Abell}, {Allison}, {Anderson}, {Andrew}, {Angel}, {Armus},
  {Arnett}, {Asztalos}, {Axelrod}, {Bailey}, {Ballantyne}, {Bankert},
  {Barkhouse}, {Barr}, {Barrientos}, {Barth}, {Bartlett}, {Becker}, {Becla},
  {Beers}, {Bernstein}, {Biswas}, {Blanton}, {Bloom}, {Bochanski}, {Boeshaar},
  {Borne}, {Bradac}, {Brandt}, {Bridge}, {Brown}, {Brunner}, {Bullock},
  {Burgasser}, {Burge}, {Burke}, {Cargile}, {Chand rasekharan}, {Chartas},
  {Chesley}, {Chu}, {Cinabro}, {Claire}, {Claver}, {Clowe}, {Connolly}, {Cook},
  {Cooke}, {Cooray}, {Covey}, {Culliton}, {de Jong}, {de Vries}, {Debattista},
  {Delgado}, {Dell'Antonio}, {Dhital}, {Di Stefano}, {Dickinson}, {Dilday},
  {Djorgovski}, {Dobler}, {Donalek}, {Dubois-Felsmann}, {Durech},
  {Eliasdottir}, {Eracleous}, {Eyer}, {Falco}, {Fan}, {Fassnacht}, {Ferguson},
  {Fernandez}, {Fields}, {Finkbeiner}, {Figueroa}, {Fox}, {Francke}, {Frank},
  {Frieman}, {Fromenteau}, {Furqan}, {Galaz}, {Gal-Yam}, {Garnavich},
  {Gawiser}, {Geary}, {Gee}, {Gibson}, {Gilmore}, {Grace}, {Green}, {Gressler},
  {Grillmair}, {Habib}, {Haggerty}, {Hamuy}, {Harris}, {Hawley}, {Heavens},
  {Hebb}, {Henry}, {Hileman}, {Hilton}, {Hoadley}, {Holberg}, {Holman},
  {Howell}, {Infante}, {Ivezic}, {Jacoby}, {Jain}, {R}, {Jedicke}, {Jee},
  {Garrett Jernigan}, {Jha}, {Johnston}, {Jones}, {Juric}, {Kaasalainen},
  {Styliani}, {Kafka}, {Kahn}, {Kaib}, {Kalirai}, {Kantor}, {Kasliwal},
  {Keeton}, {Kessler}, {Knezevic}, {Kowalski}, {Krabbendam}, {Krughoff},
  {Kulkarni}, {Kuhlman}, {Lacy}, {Lepine}, {Liang}, {Lien}, {Lira}, {Long},
  {Lorenz}, {Lotz}, {Lupton}, {Lutz}, {Macri}, {Mahabal}, {Mandelbaum},
  {Marshall}, {May}, {McGehee}, {Meadows}, {Meert}, {Milani}, {Miller},
  {Miller}, {Mills}, {Minniti}, {Monet}, {Mukadam}, {Nakar}, {Neill}, {Newman},
  {Nikolaev}, {Nordby}, {O'Connor}, {Oguri}, {Oliver}, {Olivier}, {Olsen},
  {Olsen}, {Olszewski}, {Oluseyi}, {Padilla}, {Parker}, {Pepper}, {Peterson},
  {Petry}, {Pinto}, {Pizagno}, {Popescu}, {Prsa}, {Radcka}, {Raddick},
  {Rasmussen}, {Rau}, {Rho}, {Rhoads}, {Richards}, {Ridgway}, {Robertson},
  {Roskar}, {Saha}, {Sarajedini}, {Scannapieco}, {Schalk}, {Schindler},
  {Schmidt}, {Schmidt}, {Schneider}, {Schumacher}, {Scranton}, {Sebag},
  {Seppala}, {Shemmer}, {Simon}, {Sivertz}, {Smith}, {Allyn Smith}, {Smith},
  {Spitz}, {Stanford}, {Stassun}, {Strader}, {Strauss}, {Stubbs}, {Sweeney},
  {Szalay}, {Szkody}, {Takada}, {Thorman}, {Trilling}, {Trimble}, {Tyson}, {Van
  Berg}, {Vand en Berk}, {VanderPlas}, {Verde}, {Vrsnak}, {Walkowicz}, {Wand
  elt}, {Wang}, {Wang}, {Warner}, {Wechsler}, {West}, {Wiecha}, {Williams},
  {Willman}, {Wittman}, {Wolff}, {Wood-Vasey}, {Wozniak}, {Young}, {Zentner},
  \& {Zhan}}]{lsst09}
{LSST Science Collaboration}, {Abell}, P.~A., {Allison}, J., {et~al.} 2009,
  arXiv e-prints, arXiv:0912.0201.
\newblock \doarXiv{0912.0201}

\bibitem[{{Madhavacheril} {et~al.}(2015){Madhavacheril}, {Sehgal}, {Allison},
  {Battaglia}, {Bond}, {Calabrese}, {Caligiuri}, {Coughlin}, {Crichton},
  {Datta}, {Devlin}, {Dunkley}, {D{\"u}nner}, {Fogarty}, {Grace}, {Hajian},
  {Hasselfield}, {Hill}, {Hilton}, {Hincks}, {Hlozek}, {Hughes}, {Kosowsky},
  {Louis}, {Lungu}, {McMahon}, {Moodley}, {Munson}, {Naess}, {Nati},
  {Newburgh}, {Niemack}, {Page}, {Partridge}, {Schmitt}, {Sherwin}, {Sievers},
  {Spergel}, {Staggs}, {Thornton}, {Van Engelen}, {Ward}, {Wollack}, \&
  {Atacama Cosmology Telescope Collaboration}}]{madhavacheril15}
{Madhavacheril}, M., {Sehgal}, N., {Allison}, R., {et~al.} 2015, \prl, 114,
  151302, \dodoi{10.1103/PhysRevLett.114.151302}

\bibitem[{{Mantz} {et~al.}(2008){Mantz}, {Allen}, {Ebeling}, \&
  {Rapetti}}]{mantz08}
{Mantz}, A., {Allen}, S.~W., {Ebeling}, H., \& {Rapetti}, D. 2008, \mnras, 387,
  1179, \dodoi{10.1111/j.1365-2966.2008.13311.x}

\bibitem[{{Mantz} {et~al.}(2016){Mantz}, {Allen}, {Morris}, {von der Linden},
  {Applegate}, {Kelly}, {Burke}, {Donovan}, \& {Ebeling}}]{mantz16}
{Mantz}, A.~B., {Allen}, S.~W., {Morris}, R.~G., {et~al.} 2016, \mnras, 463,
  3582, \dodoi{10.1093/mnras/stw2250}

\bibitem[{{Mathuriya} {et~al.}(2018){Mathuriya}, {Bard}, {Mendygral},
  {Meadows}, {Arnemann}, {Shao}, {He}, {Karna}, {Moise}, {Pennycook},
  {Maschoff}, {Sewall}, {Kumar}, {Ho}, {Ringenburg}, {Prabhat}, \&
  {Lee}}]{mathuriya18d}
{Mathuriya}, A., {Bard}, D., {Mendygral}, P., {et~al.} 2018, arXiv e-prints,
  arXiv:1808.04728.
\newblock \doarXiv{1808.04728}

\bibitem[{{McClintock} {et~al.}(2019){McClintock}, {Varga}, {Gruen}, {Rozo},
  {Rykoff}, {Shin}, {Melchior}, {DeRose}, {Seitz}, {Dietrich}, {Sheldon},
  {Zhang}, {von der Linden}, {Jeltema}, {Mantz}, {Romer}, {Allen}, {Becker},
  {Bermeo}, {Bhargava}, {Costanzi}, {Everett}, {Farahi}, {Hamaus}, {Hartley},
  {Hollowood}, {Hoyle}, {Israel}, {Li}, {MacCrann}, {Morris}, {Palmese},
  {Plazas}, {Pollina}, {Rau}, {Simet}, {Soares-Santos}, {Troxel}, {Vergara
  Cervantes}, {Wechsler}, {Zuntz}, {Abbott}, {Abdalla}, {Allam}, {Annis},
  {Avila}, {Bridle}, {Brooks}, {Burke}, {Carnero Rosell}, {Carrasco Kind},
  {Carretero}, {Castander}, {Crocce}, {Cunha}, {D'Andrea}, {da Costa}, {Davis},
  {De Vicente}, {Diehl}, {Doel}, {Drlica-Wagner}, {Evrard}, {Flaugher},
  {Fosalba}, {Frieman}, {Garc{\'\i}a-Bellido}, {Gaztanaga}, {Gerdes},
  {Giannantonio}, {Gruendl}, {Gutierrez}, {Honscheid}, {James}, {Kirk},
  {Krause}, {Kuehn}, {Lahav}, {Li}, {Lima}, {March}, {Marshall}, {Menanteau},
  {Miquel}, {Mohr}, {Nord}, {Ogando}, {Roodman}, {Sanchez}, {Scarpine},
  {Schindler}, {Sevilla-Noarbe}, {Smith}, {Smith}, {Sobreira}, {Suchyta},
  {Swanson}, {Tarle}, {Tucker}, {Vikram}, {Walker}, {Weller}, \& {DES
  Collaboration}}]{mclintock19}
{McClintock}, T., {Varga}, T.~N., {Gruen}, D., {et~al.} 2019, \mnras, 482,
  1352, \dodoi{10.1093/mnras/sty2711}

\bibitem[{{Monaghan} \& {Lattanzio}(1985)}]{monaghan85}
{Monaghan}, J.~J., \& {Lattanzio}, J.~C. 1985, \aap, 149, 135

\bibitem[{{Murata} {et~al.}(2019){Murata}, {Oguri}, {Nishimichi}, {Takada},
  {Mandelbaum}, {More}, {Shirasaki}, {Nishizawa}, \& {Osato}}]{murata19}
{Murata}, R., {Oguri}, M., {Nishimichi}, T., {et~al.} 2019, \pasj, 71, 107,
  \dodoi{10.1093/pasj/psz092}

\bibitem[{{Nagai} {et~al.}(2007){Nagai}, {Kravtsov}, \& {Vikhlinin}}]{nagai07}
{Nagai}, D., {Kravtsov}, A.~V., \& {Vikhlinin}, A. 2007, \apj, 668, 1,
  \dodoi{10.1086/521328}

\bibitem[{{Ntampaka} {et~al.}(2015){Ntampaka}, {Trac}, {Sutherland},
  {Battaglia}, {P{\'o}czos}, \& {Schneider}}]{ntampaka15}
{Ntampaka}, M., {Trac}, H., {Sutherland}, D.~J., {et~al.} 2015, \apj, 803, 50,
  \dodoi{10.1088/0004-637X/803/2/50}

\bibitem[{{Ntampaka} {et~al.}(2019){Ntampaka}, {ZuHone}, {Eisenstein}, {Nagai},
  {Vikhlinin}, {Hernquist}, {Marinacci}, {Nelson}, {Pakmor}, {Pillepich},
  {Torrey}, \& {Vogelsberger}}]{ntampaka19a}
{Ntampaka}, M., {ZuHone}, J., {Eisenstein}, D., {et~al.} 2019, \apj, 876, 82,
  \dodoi{10.3847/1538-4357/ab14eb}

\bibitem[{{Nwankpa} {et~al.}(2018){Nwankpa}, {Ijomah}, {Gachagan}, \&
  {Marshall}}]{nwankpa18d}
{Nwankpa}, C., {Ijomah}, W., {Gachagan}, A., \& {Marshall}, S. 2018, arXiv
  e-prints, arXiv:1811.03378.
\newblock \doarXiv{1811.03378}

\bibitem[{{Planck Collaboration} {et~al.}(2015){Planck Collaboration}, {Ade},
  {Aghanim}, {Arnaud}, {Ashdown}, {Aumont}, {Baccigalupi}, {Banday},
  {Barreiro}, {Bartlett}, \& et~al.}]{Planck15}
{Planck Collaboration}, {Ade}, P.~A.~R., {Aghanim}, N., {et~al.} 2015, ArXiv
  e-prints.
\newblock \doarXiv{1502.01597}

\bibitem[{{Planck Collaboration} {et~al.}(2016){Planck Collaboration}, {Ade},
  {Aghanim}, {Arnaud}, {Ashdown}, {Aumont}, {Baccigalupi}, {Banday},
  {Barreiro}, {Bartlett}, \& et~al.}]{planck16-24}
---. 2016, \aap, 594, A24, \dodoi{10.1051/0004-6361/201525833}

\bibitem[{{Planck Collaboration} {et~al.}(2018){Planck Collaboration},
  {Aghanim}, {Akrami}, {Ashdown}, {Aumont}, {Baccigalupi}, {Ballardini},
  {Banday}, {Barreiro}, {Bartolo}, {Basak}, {Battye}, {Benabed}, {Bernard},
  {Bersanelli}, {Bielewicz}, {Bock}, {Bond}, {Borrill}, {Bouchet}, {Boulanger},
  {Bucher}, {Burigana}, {Butler}, {Calabrese}, {Cardoso}, {Carron},
  {Challinor}, {Chiang}, {Chluba}, {Colombo}, {Combet}, {Contreras}, {Crill},
  {Cuttaia}, {de Bernardis}, {de Zotti}, {Delabrouille}, {Delouis}, {Di
  Valentino}, {Diego}, {Dor{\'e}}, {Douspis}, {Ducout}, {Dupac}, {Dusini},
  {Efstathiou}, {Elsner}, {En{\ss}lin}, {Eriksen}, {Fantaye}, {Farhang},
  {Fergusson}, {Fernandez-Cobos}, {Finelli}, {Forastieri}, {Frailis},
  {Fraisse}, {Franceschi}, {Frolov}, {Galeotta}, {Galli}, {Ganga},
  {G{\'e}nova-Santos}, {Gerbino}, {Ghosh}, {Gonz{\'a}lez-Nuevo}, {G{\'o}rski},
  {Gratton}, {Gruppuso}, {Gudmundsson}, {Hamann}, {Handley}, {Hansen},
  {Herranz}, {Hildebrandt}, {Hivon}, {Huang}, {Jaffe}, {Jones}, {Karakci},
  {Keih{\"a}nen}, {Keskitalo}, {Kiiveri}, {Kim}, {Kisner}, {Knox},
  {Krachmalnicoff}, {Kunz}, {Kurki-Suonio}, {Lagache}, {Lamarre}, {Lasenby},
  {Lattanzi}, {Lawrence}, {Le Jeune}, {Lemos}, {Lesgourgues}, {Levrier},
  {Lewis}, {Liguori}, {Lilje}, {Lilley}, {Lindholm}, {L{\'o}pez-Caniego},
  {Lubin}, {Ma}, {Mac{\'\i}as-P{\'e}rez}, {Maggio}, {Maino}, {Mandolesi},
  {Mangilli}, {Marcos-Caballero}, {Maris}, {Martin}, {Martinelli},
  {Mart{\'\i}nez-Gonz{\'a}lez}, {Matarrese}, {Mauri}, {McEwen}, {Meinhold},
  {Melchiorri}, {Mennella}, {Migliaccio}, {Millea}, {Mitra},
  {Miville-Desch{\^e}nes}, {Molinari}, {Montier}, {Morgante}, {Moss}, {Natoli},
  {N{\o}rgaard-Nielsen}, {Pagano}, {Paoletti}, {Partridge}, {Patanchon},
  {Peiris}, {Perrotta}, {Pettorino}, {Piacentini}, {Polastri}, {Polenta},
  {Puget}, {Rachen}, {Reinecke}, {Remazeilles}, {Renzi}, {Rocha}, {Rosset},
  {Roudier}, {Rubi{\~n}o-Mart{\'\i}n}, {Ruiz-Granados}, {Salvati}, {Sandri},
  {Savelainen}, {Scott}, {Shellard}, {Sirignano}, {Sirri}, {Spencer},
  {Sunyaev}, {Suur-Uski}, {Tauber}, {Tavagnacco}, {Tenti}, {Toffolatti},
  {Tomasi}, {Trombetti}, {Valenziano}, {Valiviita}, {Van Tent}, {Vibert},
  {Vielva}, {Villa}, {Vittorio}, {Wand elt}, {Wehus}, {White}, {White},
  {Zacchei}, \& {Zonca}}]{planck18-1}
{Planck Collaboration}, {Aghanim}, N., {Akrami}, Y., {et~al.} 2018, arXiv
  e-prints, arXiv:1807.06209.
\newblock \doarXiv{1807.06209}

\bibitem[{{Raghunathan} {et~al.}(2019){Raghunathan}, {Patil}, {Baxter},
  {Benson}, {Bleem}, {Chou}, {Crawford}, {Holder}, {McClintock}, {Reichardt},
  {Rozo}, {Varga}, {Abbott}, {Ade}, {Allam}, {Anderson}, {Annis}, {Austermann},
  {Avila}, {Beall}, {Bechtol}, {Bender}, {Bernstein}, {Bertin}, {Bianchini},
  {Brooks}, {Burke}, {Carlstrom}, {Carretero}, {Chang}, {Chiang}, {Cho},
  {Citron}, {Crites}, {Cunha}, {da Costa}, {Davis}, {Desai}, {Diehl},
  {Dietrich}, {Dobbs}, {Doel}, {Eifler}, {Everett}, {Evrard}, {Flaugher},
  {Fosalba}, {Frieman}, {Gallicchio}, {Garc{\'\i}a-Bellido}, {Gaztanaga},
  {George}, {Gilbert}, {Gruen}, {Gruendl}, {Gschwend}, {Gupta}, {Gutierrez},
  {de Haan}, {Halverson}, {Harrington}, {Hartley}, {Henning}, {Hilton},
  {Hollowood}, {Holzapfel}, {Honscheid}, {Hou}, {Hoyle}, {Hrubes}, {Huang},
  {Hubmayr}, {Irwin}, {James}, {Jeltema}, {Kim}, {Carrasco Kind}, {Knox},
  {Kovacs}, {Kuehn}, {Kuropatkin}, {Lee}, {Li}, {Lima}, {Maia}, {Marshall},
  {McMahon}, {Melchior}, {Menanteau}, {Meyer}, {Miller}, {Miquel}, {Mocanu},
  {Montgomery}, {Nadolski}, {Natoli}, {Nibarger}, {Novosad}, {Padin}, {Plazas},
  {Pryke}, {Rapetti}, {Romer}, {Carnero Rosell}, {Ruhl}, {Saliwanchik},
  {Sanchez}, {Sayre}, {Scarpine}, {Schaffer}, {Schubnell}, {Serrano},
  {Sevilla-Noarbe}, {Smecher}, {Smith}, {Soares-Santos}, {Sobreira}, {Stark},
  {Story}, {Suchyta}, {Swanson}, {Tarle}, {Thomas}, {Tucker}, {Vanderlinde},
  {De Vicente}, {Vieira}, {Wang}, {Whitehorn}, {Wu}, \&
  {Zhang}}]{raghunathan19}
{Raghunathan}, S., {Patil}, S., {Baxter}, E., {et~al.} 2019, \apj, 872, 170,
  \dodoi{10.3847/1538-4357/ab01ca}

\bibitem[{{Ronneberger} {et~al.}(2015){Ronneberger}, {Fischer}, \&
  {Brox}}]{ronneberger15d}
{Ronneberger}, O., {Fischer}, P., \& {Brox}, T. 2015, arXiv e-prints,
  arXiv:1505.04597.
\newblock \doarXiv{1505.04597}

\bibitem[{{Ruder}(2016)}]{ruder16d}
{Ruder}, S. 2016, arXiv e-prints, arXiv:1609.04747.
\newblock \doarXiv{1609.04747}

\bibitem[{{Sif{\'o}n} {et~al.}(2013){Sif{\'o}n}, {Menanteau}, {Hasselfield},
  {Marriage}, {Hughes}, {Barrientos}, {Gonz{\'a}lez}, {Infante}, {Addison},
  {Baker}, {Battaglia}, {Bond}, {Crichton}, {Das}, {Devlin}, {Dunkley},
  {D{\"u}nner}, {Gralla}, {Hajian}, {Hilton}, {Hincks}, {Kosowsky}, {Marsden},
  {Moodley}, {Niemack}, {Nolta}, {Page}, {Partridge}, {Reese}, {Sehgal},
  {Sievers}, {Spergel}, {Staggs}, {Thornton}, {Trac}, \& {Wollack}}]{sifon13}
{Sif{\'o}n}, C., {Menanteau}, F., {Hasselfield}, M., {et~al.} 2013, \apj, 772,
  25, \dodoi{10.1088/0004-637X/772/1/25}

\bibitem[{{Sif{\'o}n} {et~al.}(2016){Sif{\'o}n}, {Battaglia}, {Hasselfield},
  {Menanteau}, {Barrientos}, {Bond}, {Crichton}, {Devlin}, {D{\"u}nner},
  {Hilton}, {Hincks}, {Hlozek}, {Huffenberger}, {Hughes}, {Infante},
  {Kosowsky}, {Marsden}, {Marriage}, {Moodley}, {Niemack}, {Page}, {Spergel},
  {Staggs}, {Trac}, \& {Wollack}}]{sifon16}
{Sif{\'o}n}, C., {Battaglia}, N., {Hasselfield}, M., {et~al.} 2016, \mnras,
  461, 248, \dodoi{10.1093/mnras/stw1284}

\bibitem[{{Soergel} {et~al.}(2018){Soergel}, {Saro}, {Giannantonio},
  {Efstathiou}, \& {Dolag}}]{soergel18}
{Soergel}, B., {Saro}, A., {Giannantonio}, T., {Efstathiou}, G., \& {Dolag}, K.
  2018, \mnras, 478, 5320, \dodoi{10.1093/mnras/sty1324}

\bibitem[{{Staniszewski} {et~al.}(2009){Staniszewski}, {Ade}, {Aird}, {Benson},
  {Bleem}, {Carlstrom}, {Chang}, {Cho}, {Crawford}, {Crites}, {de Haan},
  {Dobbs}, {Halverson}, {Holder}, {Holzapfel}, {Hrubes}, {Joy}, {Keisler},
  {Lanting}, {Lee}, {Leitch}, {Loehr}, {Lueker}, {McMahon}, {Mehl}, {Meyer},
  {Mohr}, {Montroy}, {Ngeow}, {Padin}, {Plagge}, {Pryke}, {Reichardt}, {Ruhl},
  {Schaffer}, {Shaw}, {Shirokoff}, {Spieler}, {Stalder}, {Stark},
  {Vanderlinde}, {Vieira}, {Zahn}, \& {Zenteno}}]{staniszewski09}
{Staniszewski}, Z., {Ade}, P.~A.~R., {Aird}, K.~A., {et~al.} 2009, \apj, 701,
  32, \dodoi{10.1088/0004-637X/701/1/32}

\bibitem[{{Stern} {et~al.}(2019){Stern}, {Dietrich}, {Bocquet}, {Applegate},
  {Mohr}, {Bridle}, {Carrasco Kind}, {Gruen}, {Jarvis}, {Kacprzak}, {Saro},
  {Sheldon}, {Troxel}, {Zuntz}, {Benson}, {Capasso}, {Chiu}, {Desai},
  {Rapetti}, {Reichardt}, {Saliwanchik}, {Schrabback}, {Gupta}, {Abbott},
  {Abdalla}, {Avila}, {Bertin}, {Brooks}, {Burke}, {Carnero Rosell},
  {Carretero}, {Castander}, {D'Andrea}, {da Costa}, {Davis}, {De Vicente},
  {Diehl}, {Doel}, {Estrada}, {Evrard}, {Flaugher}, {Fosalba}, {Frieman},
  {Garc{\'\i}a-Bellido}, {Gaztanaga}, {Gruendl}, {Gschwend}, {Gutierrez},
  {Hollowood}, {Jeltema}, {Kirk}, {Kuehn}, {Kuropatkin}, {Lahav}, {Lima},
  {Maia}, {March}, {Melchior}, {Menanteau}, {Miquel}, {Plazas}, {Romer},
  {Sanchez}, {Schindler}, {Schubnell}, {Sevilla-Noarbe}, {Smith}, {Smith},
  {Sobreira}, {Suchyta}, {Swanson}, {Tarle}, {Walker}, {DES Collaboration}, \&
  {SPT Collaboration}}]{stern18}
{Stern}, C., {Dietrich}, J.~P., {Bocquet}, S., {et~al.} 2019, \mnras, 485, 69,
  \dodoi{10.1093/mnras/stz234}

\bibitem[{{Sunyaev} \& {Zel'dovich}(1970)}]{sunyaev70}
{Sunyaev}, R.~A., \& {Zel'dovich}, Y.~B. 1970, Comments on Astrophysics and
  Space Physics, 2, 66

\bibitem[{{Sunyaev} \& {Zel'dovich}(1972)}]{sunyaev72}
---. 1972, Comments on Astrophysics and Space Physics, 4, 173

\bibitem[{{The Planck Collaboration}(2006)}]{planck06}
{The Planck Collaboration}. 2006, ArXiv:astro-ph/0604069

\bibitem[{{Vikhlinin} {et~al.}(2009){Vikhlinin}, {Kravtsov}, {Burenin},
  {Ebeling}, {Forman}, {Hornstrup}, {Jones}, {Murray}, {Nagai}, {Quintana}, \&
  {Voevodkin}}]{vikhlinin09}
{Vikhlinin}, A., {Kravtsov}, A.~V., {Burenin}, R.~A., {et~al.} 2009, \apj, 692,
  1060, \dodoi{10.1088/0004-637X/692/2/1060}

\bibitem[{{Yu} \& {Koltun}(2015)}]{yu15d}
{Yu}, F., \& {Koltun}, V. 2015, arXiv e-prints, arXiv:1511.07122.
\newblock \doarXiv{1511.07122}

\bibitem[{{Zhang} {et~al.}(2018){Zhang}, {Liu}, \& {Wang}}]{zhang18d}
{Zhang}, Z., {Liu}, Q., \& {Wang}, Y. 2018, IEEE Geoscience and Remote Sensing
  Letters, 15, 749, \dodoi{10.1109/LGRS.2018.2802944}

\end{thebibliography}

\end{document}